\documentclass{aa} 
\usepackage{graphicx,epsfig,times}

\newcommand{\chandra}{{\it Chandra} }
\newcommand{\asca}{{\it ASCA} }
\newcommand{\sax}{{\it BeppoSAX} }
\newcommand{\rosat}{{\it ROSAT} }

\title{Scaling laws in X-ray Galaxy Clusters \\ at redshift between 0.4 and 1.3}
\titlerunning{Scaling laws in high-$z$ X-ray Galaxy Clusters}

\author{S. Ettori\inst{1} \and P. Tozzi\inst{2} \and 
S. Borgani\inst{3,4} \and P. Rosati\inst{1} } 
\authorrunning{S. Ettori et al.}

\institute{
 ESO, Karl-Schwarzschild-Str. 2, D-85748 Garching, Germany
 \and INAF, Osservatorio Astronomico di Trieste, via G.B. Tiepolo 11, I-34131 Trieste, Italy
 \and Dip. di Astronomia, Universit\'a di Trieste, via G.B. Tiepolo 11, I-34131 Trieste, Italy
 \and INFN -- Istituto Nazionale di Fisica Nucleare, Trieste, Italy
}

\offprints{S. Ettori}

\mail{settori@eso.org}

\date{Submitted on 25 July 2003, accepted on 4 December 2003}

\begin{document}

\abstract{ We present a study of the integrated physical properties of
a sample of 28 X-ray galaxy clusters observed with \chandra at
a redshift between 0.4 and 1.3. In particular, we have twelve objects in
the redshift range 0.4--0.6, five between 0.6 and 0.8, seven between
0.8 and 1 and four at $z>$1.0, compounding the largest sample
available for such a study.  We focus particularly on the properties
and evolution of the X-ray scaling laws. 
We fit both a single and 
a double $\beta-$model with the former which provides a good representation
of the observed surface brightness profiles, indicating that these
clusters do not show any significant excess in their central brightness.
By using the best-fit parameters of the $\beta-$model together with
the measured emission-weighted temperature (in the range 3--11 keV), 
we recover gas luminosity, gas mass and total gravitating mass
out to $R_{500}$. We observe scaling relations steeper than expected from
the self-similar model by a significant ($>3 \sigma$) amount in the $L-T$
and $M_{\rm gas}-T$ relations and by a marginal value in the $M_{\rm
tot}-T$ and $L-M_{\rm tot}$ relations. The degree of evolution of the
$M_{\rm tot}-T$ relation is found to be consistent with the
expectation based on the hydrostatic equilibrium for gas within
virialized dark matter halos.  We detect hints of {\it negative}
evolution in the $L-T$, $M_{\rm gas}-T$ and $L-M_{\rm tot}$ relations, 
thus suggesting that systems at higher redshift have lower X-ray 
luminosity and gas mass for fixed temperature. 
In particular, 
when the 16 clusters at $z>0.6$ are considered, the evolution becomes
more evident and its power-law modelization is a statistically 
good description of the data.
In this subsample, we also find significant evidence for 
{\it positive} evolution, such as $(1+z)^{0.3}$, in the $E_z^{4/3} S - T$
relation, where the entropy $S$ is defined as $T/n_{\rm gas}^{2/3}$
and is measured at 0.1$\,R_{200}$. 
Such results point toward a scenario in which a relatively lower gas
density is present in high--redshift objects, thus implying a
suppressed X--ray emission, a smaller amount of gas mass and a higher
entropy level. This represents a non--trivial constraint for models
aiming at explaining the thermal history of the intra--cluster medium
out to the highest redshift reached so far.
}

\maketitle


\section{Introduction}

The physics of the intracluster medium (ICM) is mainly driven by the
infall of the cosmic baryons trapped in the deep gravitational
potential of the cluster dark matter halo. Through a hierarchical
formation that starts from the primordial density fluctuations and
generates the largest virialized structures via gravitational collapse
and merging, galaxy clusters maintain similar properties when they are
rescaled with respect to gravitational mass and epoch of
formation. The shock--heated X-ray emitting ICM accounts for most of the 
baryons collapsed in the cluster potential (e.g. White et al. 1993).
Its physical properties, like density and temperature,
relate in a predictable way in this simple self-similar scenario
(e.g. Kaiser 1986, 1991, Evrard \& Henry 1991).  Under the assumptions
that the smooth and spherically symmetric ICM is heated only by
gravitational processes (adiabatic compression during DM collapse and
shock heating from supersonic accretion), obeys hydrostatic
equilibrium within the underlying dark matter potential and emits
mainly by bremsstrahlung, one derives the following scaling
relations between the observed (bolometric luminosity $L_{\rm bol}$,
and temperature $T_{\rm gas}$) and derived (gas entropy $S$, gas mass
$M_{\rm gas}$, and total gravitating mass $M_{\rm tot}$) quantities:
\begin{itemize}
\item $E_z^{4/3} \; S \propto \Delta_z^{-2/3} \; T_{\rm gas}$
\item $E_z^{-1} \; L_{\rm bol} \; \propto \; \Delta_z^{1/2} \; T_{\rm gas}^2$
\item $E_z \; M_{\rm tot} \; \propto \; \Delta_z^{-1/2} \; T_{\rm gas}^{3/2}$
\item $E_z^{-1} \; L_{\rm bol} \; \propto \; \Delta_z^{7/6} \; (E_z
  M_{\rm tot})^{4/3}$
\item $E_z \; M_{\rm gas} \; \propto \; \Delta_z^{-1/2} \; T_{\rm gas}^{3/2}$.
\end{itemize}
The factors that indicate the dependence on the evolution of the
Hubble constant at redshift $z$, $E_z = H_z / H_0 = \left[\Omega_{\rm
m} (1+z)^3 + 1 - \Omega_{\rm m} \right]^{1/2}$ (for a flat cosmology
with matter density $\Omega_{\rm m}$), and on the overdensity $\Delta_z$
account for the fact that all the quantities are estimated 
at a given overdensity $\Delta_z$ with respect to the critical density
estimated at redshift $z$,
$\rho_{\rm c, z} = 3 H_z^2/ (8 \pi G)$.

\begin{figure}
 \epsfig{figure=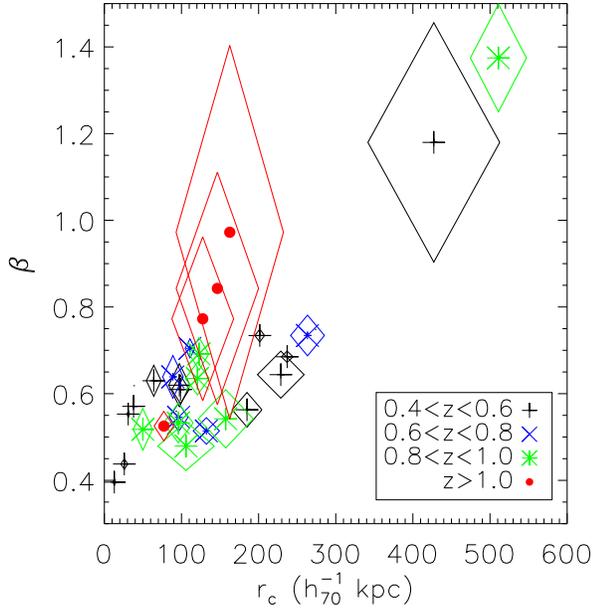,width=0.5\textwidth}
\caption{Correlations between the best-fit parameters of 
the $\beta-$model: higher values of the core radius, $r_{\rm c}$, correspond
to higher estimates of $\beta$.  No redshift-segregation is evident.
} \label{fig:bmod_par} \end{figure} 

Hydrodynamical simulations (e.g. Evrard, Metzler, Navarro 1996, Bryan
\& Norman 1998, Thomas et al. 2001, Bialek et al. 2001, Borgani et
al. 2002 and references therein) and observational analyses (e.g. from
Mushotzky 1984, Edge \& Stewart 1991 to the more recent work of
Allen \& Fabian 1998, Markevitch 1998, Arnaud \& Evrard 1999,
Nevalainen et al. 2000, Finoguenov et al. 2001, Ettori et al. 2002) of
the best-studied correlations between X-ray luminosity, total
gravitating mass and gas temperature 
%
show that significant deviations exist between observations and the
expectations based on self--similar scaling:
(i) a difference of the order of
30--40 per cent in the normalization of the $M_{\rm tot} - T_{\rm
gas}$ relation (Horner et al. 1999, Nevalainen et al. 2000); (ii)
steeper slopes of 
the $L_{\rm bol}$--$T$, $M_{\rm gas}$--$T$ and, possibly, $M_{\rm
tot}$--$T$ relations, in particular when low temperature ($T_{\rm gas}
<$ 3 keV) systems are also considered; (iii) excess entropy in the
central regions of clusters and groups, with respect to the $S\propto
T$ expectation (e.g., Ponman et al. 1999, 2003). Such deviations are
currently interpreted as evidence for
%
%
non-gravitational processes, such as
extra heating and radiative cooling, that affect the assembly of the
baryons in the cluster potential well (e.g. Cavaliere, Menci \& Tozzi
1999, Bower et al. 2001, Tozzi \& Norman 2001, Babul et al. 2002, Voit
et al. 2002).
 
The same processes that determine the shape of the local relations
should have affected also the evolution of the scaling laws with the
cosmic epoch. However, due to the objective difficulty to assemble a
large dataset of high redshift objects, there has been very little
work done until now, and with contradictory results, on the observed
evolution of the scaling laws. For example, for different subsamples 
of our dataset, Holden et al. (2002) 
observe no evolution of the $L_{\rm bol} - T_{\rm gas}$ relation in a
$\Lambda$CDM universe, while Vikhlinin et al. (2002) claim to have
evidence at $8 \sigma$ level for a positive evolution, i.e. systems at
higher redshift have higher luminosity at fixed temperature.  In
principle, any observed evolution in the scaling laws allows to
constrain different scenarios of heating schemes and cooling
efficiency.

Furthermore, understanding the evolution of scaling relations
involving luminosity, temperature and total collapsed mass, is of
vital importance if the evolution of the cluster population has to be
used as a proxy to the cosmic evolution and, therefore, for the
determination of cosmological parameters (e.g., the pioneering work of
Henry \& Arnaud 1991; see Rosati, Borgani \& Norman 2002, for a recent
review and references therein). In fact, while the mass function of
galaxy clusters and its evolution are sensitive probes of the
underlying cosmological scenario, mass is usually inferred from
observational quantities, such as the X--ray luminosity and the gas
temperature, which are affected by non-gravitational processes.

In this paper, we focus on the general behaviour of the scaling laws
for X-ray clusters at high redshift, with the main aim of quantifying 
their evolution with respect to what is observed for nearby clusters. 
To do this, we have assembled
a set of 28 clusters in the redshift range 0.4--1.3 observed at arcsec 
angular resolution with \chandra (Weisskopf et al. 2000).  
This allows us to resolve on scales of few
tens of kpc the surface brightness (i.e. the gas density) of these
systems, and to determine a single emission weighted temperature.  
In the next section, we describe our dataset and how the physical
quantities are measured from the observations.  We present in
Section~3 the results on the scaling laws under examination and in 
Section~4 the constraints on the observed evolution of these relations. 
We summarize our conclusions in Section~5.
All quoted errors are at $1 \sigma$ level (68.3 per cent level of confidence 
and $\Delta \chi^2=1$ for one interesting parameter) unless
otherwise stated. The cosmological parameters $H_0 = 70 h^{-1}_{70}$
km s$^{-1}$ Mpc$^{-1}$ and $\Omega_{\rm m} = 1 -\Omega_{\Lambda} =
0.3$ are assumed hereafter as suggested from WMAP constraints on the
anisotropies in the cosmic microwave background (Spergel et al. 2003).

\begin{table*} 
\begin{center}
\caption{Best-fit results of the spectral and spatial analysis
of the sample of galaxy clusters at redshift $>0.4$ and derived
quantities at an overdensity corresponding to $\Delta_z=500$
in a Einstein-de Sitter universe.
A Hubble constant of 70 km s$^{-1}$ Mpc$^{-1}$ in a flat universe
with $\Omega_{\rm m}$ equals to 0.3 is assumed.
Note: (i) RXJ1347--1145 has a clear hot enhancement to the South-East of the
main X-ray emission. A sector with position angle $90^{\circ}-180^{\circ}$ 
and centered
on the X-ray peak has been masked in accordance with the analysis presented in 
Allen, Schmidt \& Fabian (2002). (ii) In 3C295, the central active galactic 
nucleus and regions of enhanced emission associated to the radio lobes 
have been masked (see Allen et al. 2001a). 
(iii) MS1054--0321 presents significant substructure (e.g. Jeltema et al. 
2001). The temperature and the best-fit of the surface brightness profile are 
estimated from the main body of the cluster once a circular region centered at 
(RA, Dec; 2000)=$(10^{\rm h} 56^{\rm m} 55"7,-3^o 37' 37'')$ 
and with radius of 36 arcsec is masked.
}
\begin{tabular}{l@{\hspace{.7em}} c@{\hspace{.7em}} c@{\hspace{.7em}} c@{\hspace{.7em}} c@{\hspace{.7em}} c@{\hspace{.7em}} c@{\hspace{.7em}} c@{\hspace{.7em}} c@{\hspace{.7em}} c@{\hspace{.7em}} c@{\hspace{.7em}} c@{\hspace{.7em}}}
\hline \\ 
cluster & $z$ & $1'$ & $R_{\rm spec}$ & $R_{2 \sigma}$ & $T_{\rm gas}$ & 
$r_{\rm c}$ & $\beta$ & $R_{500}$ & $L_{\rm bol}$ & $M_{\rm gas}$ & $M_{\rm tot}$ \\ 
  & & kpc & kpc & kpc & keV & kpc &  & kpc & $10^{44}$ erg s$^{-1}$ & $10^{13} M_{\odot}$ & $10^{14} M_{\odot}$  \\ 
  & & & & & & & & & & &\\ 
\hline \\
RXJ1416+4446 & 0.400 & $322$ & $397$ & $1088$ & $3.7_{-0.3}^{+0.2}$ & $26(\pm 4)$ & $0.438(\pm 0.012)$ & $749(\pm 26)$ & $5.43(\pm 0.19)$ & $2.84(\pm 0.47)$ & $1.35(\pm 0.14)$ \\ 
MS0302+1658 & 0.424 & $334$ & $274$ & $552$ & $3.8_{-0.2}^{+0.9}$ & $64(\pm 10)$ & $0.630(\pm 0.036)$ & $899(\pm 65)$ & $6.41(\pm 0.19)$ & $3.26(\pm 0.48)$ & $2.43(\pm 0.55)$ \\ 
MS1621+2640 & 0.426 & $335$ & $659$ & $1276$ & $6.8_{-0.5}^{+0.9}$ & $185(\pm 19)$ & $0.563(\pm 0.040)$ & $1119(\pm 74)$ & $10.92(\pm 0.30)$ & $7.18(\pm 0.76)$ & $4.70(\pm 0.94)$ \\ 
RXJ1347--1145 & 0.451 & $346$ & $533$ & $1430$ & $10.3_{-0.5}^{+0.6}$ & $38(\pm 1)$ & $0.571(\pm 0.004)$ & $1368(\pm 40)$ & $116.75(\pm 0.58)$ & $18.06(\pm 0.80)$ & $8.94(\pm 0.80)$ \\ 
RXJ1701+6421 & 0.453 & $347$ & $370$ & $889$ & $4.5_{-0.2}^{+0.4}$ & $13(\pm 2)$ & $0.396(\pm 0.007)$ & $750(\pm 24)$ & $6.42(\pm 0.58)$ & $3.36(\pm 0.45)$ & $1.48(\pm 0.15)$ \\ 
3C295 & 0.460 & $350$ & $172$ & $895$ & $4.3_{-0.2}^{+0.3}$ & $31(\pm 2)$ & $0.553(\pm 0.007)$ & $870(\pm 26)$ & $14.07(\pm 0.19)$ & $4.35(\pm 0.29)$ & $2.33(\pm 0.21)$ \\ 
RXJ1525+0957 & 0.516 & $372$ & $372$ & $973$ & $5.1_{-0.5}^{+0.5}$ & $229(\pm 30)$ & $0.644(\pm 0.054)$ & $941(\pm 60)$ & $6.92(\pm 0.19)$ & $4.74(\pm 0.62)$ & $3.23(\pm 0.62)$ \\ 
MS0451--0305 & 0.539 & $381$ & $625$ & $1391$ & $8.0_{-0.3}^{+0.3}$ & $201(\pm 6)$ & $0.734(\pm 0.015)$ & $1263(\pm 30)$ & $50.94(\pm 8.05)$ & $14.15(\pm 0.47)$ & $8.09(\pm 0.59)$ \\ 
MS0016+1609 & 0.541 & $382$ & $626$ & $1770$ & $10.0_{-0.5}^{+0.5}$ & $237(\pm 8)$ & $0.685(\pm 0.013)$ & $1355(\pm 35)$ & $53.27(\pm 7.33)$ & $17.13(\pm 0.66)$ & $10.02(\pm 0.78)$ \\ 
RXJ1121+2326 & 0.562 & $389$ & $446$ & $702$ & $4.6_{-0.3}^{+0.5}$ & $427(\pm 85)$ & $1.180(\pm 0.276)$ & $1132(\pm 128)$ & $5.45(\pm 0.22)$ & $4.60(\pm 0.39)$ & $6.04(\pm 2.12)$ \\ 
RXJ0848+4456 & 0.570 & $391$ & $196$ & $401$ & $3.2_{-0.3}^{+0.3}$ & $97(\pm 14)$ & $0.620(\pm 0.050)$ & $723(\pm 45)$ & $1.21(\pm 0.37)$ & $1.23(\pm 0.14)$ & $1.59(\pm 0.30)$ \\ 
MS2053--0449 & 0.583 & $396$ & $357$ & $730$ & $5.5_{-0.5}^{+0.5}$ & $99(\pm 11)$ & $0.610(\pm 0.033)$ & $931(\pm 47)$ & $5.40(\pm 1.04)$ & $3.41(\pm 0.35)$ & $3.47(\pm 0.52)$ \\ 
RXJ0542--4100 & 0.634 & $411$ & $540$ & $1198$ & $7.9_{-0.8}^{+1.0}$ & $132(\pm 17)$ & $0.514(\pm 0.031)$ & $982(\pm 63)$ & $12.15(\pm 1.36)$ & $6.27(\pm 0.82)$ & $4.41(\pm 0.85)$ \\ 
RXJ1221+4918 & 0.700 & $429$ & $563$ & $972$ & $7.5_{-0.6}^{+0.7}$ & $263(\pm 22)$ & $0.734(\pm 0.047)$ & $1063(\pm 57)$ & $12.95(\pm 0.39)$ & $6.80(\pm 0.59)$ & $6.16(\pm 1.00)$ \\ 
RXJ1113--2615 & 0.730 & $436$ & $286$ & $371$ & $5.6_{-0.6}^{+0.8}$ & $89(\pm 13)$ & $0.639(\pm 0.049)$ & $866(\pm 62)$ & $4.43(\pm 0.77)$ & $2.49(\pm 0.31)$ & $3.47(\pm 0.75)$ \\ 
RXJ2302+0844 & 0.734 & $437$ & $358$ & $635$ & $6.6_{-0.6}^{+1.5}$ & $96(\pm 12)$ & $0.546(\pm 0.033)$ & $865(\pm 65)$ & $5.45(\pm 0.17)$ & $3.38(\pm 0.48)$ & $3.48(\pm 0.82)$ \\ 
MS1137+6625 & 0.782 & $447$ & $367$ & $635$ & $6.9_{-0.5}^{+0.5}$ & $111(\pm 6)$ & $0.705(\pm 0.022)$ & $964(\pm 39)$ & $15.30(\pm 0.44)$ & $4.90(\pm 0.29)$ & $5.18(\pm 0.63)$ \\ 
RXJ1317+2911 & 0.805 & $451$ & $185$ & $238$ & $4.1_{-0.8}^{+1.2}$ & $50(\pm 12)$ & $0.518(\pm 0.048)$ & $634(\pm 83)$ & $1.12(\pm 0.08)$ & $0.98(\pm 0.29)$ & $1.52(\pm 0.61)$ \\ 
RXJ1350+6007 & 0.810 & $452$ & $519$ & $801$ & $4.6_{-0.9}^{+0.7}$ & $106(\pm 37)$ & $0.479(\pm 0.056)$ & $627(\pm 65)$ & $4.41(\pm 0.39)$ & $2.24(\pm 0.79)$ & $1.48(\pm 0.46)$ \\ 
RXJ1716+6708 & 0.813 & $453$ & $371$ & $640$ & $6.8_{-0.6}^{+1.0}$ & $121(\pm 16)$ & $0.635(\pm 0.038)$ & $896(\pm 56)$ & $13.86(\pm 1.04)$ & $4.87(\pm 0.61)$ & $4.35(\pm 0.83)$ \\ 
MS1054--0321 & 0.830 & $456$ & $628$ & $900$ & $10.2_{-0.8}^{+1.0}$ & $511(\pm 36)$ & $1.375(\pm 0.124)$ & $1509(\pm 93)$ & $28.48(\pm 2.96)$ & $10.61(\pm 0.44)$ & $21.27(\pm 3.96)$ \\ 
RXJ0152--1357S & 0.830 & $456$ & $400$ & $607$ & $6.9_{-0.8}^{+2.9}$ & $96(\pm 19)$ & $0.532(\pm 0.046)$ & $824(\pm 109)$ & $7.73(\pm 0.40)$ & $3.91(\pm 0.94)$ & $3.47(\pm 1.49)$ \\ 
RXJ0152--1357N & 0.835 & $457$ & $427$ & $919$ & $6.0_{-0.7}^{+1.1}$ & $157(\pm 36)$ & $0.542(\pm 0.067)$ & $753(\pm 72)$ & $10.67(\pm 0.67)$ & $4.49(\pm 1.01)$ & $2.66(\pm 0.77)$ \\ 
WGA1226+3333 & 0.890 & $466$ & $459$ & $711$ & $11.2_{-1.5}^{+2.2}$ & $123(\pm 13)$ & $0.692(\pm 0.037)$ & $1134(\pm 95)$ & $54.63(\pm 0.83)$ & $10.54(\pm 1.33)$ & $9.80(\pm 2.55)$ \\ 
RXJ0910+5422 & 1.106 & $491$ & $201$ & $273$ & $6.6_{-1.3}^{+1.7}$ & $147(\pm 53)$ & $0.843(\pm 0.268)$ & $818(\pm 150)$ & $2.83(\pm 0.35)$ & $1.56(\pm 0.30)$ & $4.91(\pm 2.93)$ \\ 
RXJ1252--2927 & 1.235 & $500$ & $492$ & $440$ & $5.2_{-0.7}^{+0.7}$ & $77(\pm 13)$ & $0.525(\pm 0.034)$ & $532(\pm 40)$ & $5.99(\pm 1.10)$ & $1.81(\pm 0.34)$ & $1.59(\pm 0.35)$ \\ 
RXJ0849+4452 & 1.260 & $501$ & $197$ & $275$ & $5.2_{-1.1}^{+1.6}$ & $128(\pm 40)$ & $0.773(\pm 0.189)$ & $640(\pm 106)$ & $2.83(\pm 0.17)$ & $1.38(\pm 0.30)$ & $2.85(\pm 1.48)$ \\ 
RXJ0848+4453 & 1.270 & $501$ & $164$ & $173$ & $2.9_{-0.8}^{+0.8}$ & $163(\pm 70)$ & $0.972(\pm 0.431)$ & $499(\pm 115)$ & $1.04(\pm 0.73)$ & $0.56(\pm 0.21)$ & $1.37(\pm 0.98)$ \\ 
\hline \\ 
\end{tabular}

\end{center}
\label{tab:obj_a}
\end{table*}


\section{The dataset}

As of Summer 2003, we consider all the clusters available in the \chandra 
archive with redshift larger than 0.4.
These 28 galaxy clusters have redshift between 0.4 and 1.3 (median $z=$ 0.73), 
emission-weighted temperature in the range 3--11 keV (median value of 6.0 keV) 
and luminosity between $10^{44}$ erg s$^{-1}$ and $1.2 \times 10^{46}$ 
erg s$^{-1}$ (median value: $6.9 \times 10^{44}$ erg s$^{-1}$).  
In particular, twelve of these objects were selected from the \rosat Deep 
Cluster Survey (RDCS; Rosati et al. 1998, Stanford et al. 2001, Holden et al. 
2002, Rosati et al. 2003) of which we present the complete \chandra
follow-up of objects with $z>0.8$, 
seven from the {\it Einstein} Extended Medium Sensitivity Survey (MS; Gioia
et al. 1990), four from the 160 Square Degrees \rosat Survey
(Vikhlinin et al. 1998, Mullis et al. 2003), RXJ1113 and WGA1226 from 
the Wide Angle \rosat Pointed Survey (WARPS; Perlman et al. 2002, Maughan et
al. 2003; see also Cagnoni et al. 2001), RXJ1716 is part of
the North Ecliptic Pole survey (NEP; Gioia et al. 1999, Henry et
al. 2001), RXJ1347 has been discovered in the \rosat All Sky
Survey (Schindler et al. 1995) and 3C295 has been already mapped with 
{\it Einstein} (Henry \& Henriksen 1986).
Subsamples of the present one have been used to constrain the cosmological
parameters through the cluster baryonic content (Ettori et al. 2003) and
the Iron abundance in the ICM at high redshift (Tozzi et al. 2003).

The data reduction is described in details in Tozzi et al. (2003).  We
summarize here its most relevant aspects.  We reprocess the level=1
events files retrieved from the archive. 
When the observations were in VFAINT mode,
the pulse height in a $5\times5$ event island has been used. The
corrections for Charge Transfer Inefficiency (CTI) have been applied
when the Focal Plane temperature was 153 K.
Finally, we apply a double correction to the ARF files. 
We apply the script {\tt apply\_acisabs} by Chartas and Getman to take into
account the degradation in the ACIS QE due to material accumulated on the ACIS
optical blocking filter since launch (see {\tt http://cxc.harvard.edu/ ciao/ 
threads/ applyacisabs/} and Marshall et al. 2003).
This correction apply to all the observation.  
For data taken with ACIS-I (note that only RXJ1347, 3C295, MS0451, MS1054 
and WGA1226 have been observed with the back-illuminated S3 CCD), 
we also apply manually a correction to the effective area consisting in a 
7\% decrement below 1.8 keV (see Markevitch \& Vikhlinin 2001).

\subsection{X-ray analysis}

\begin{table}
\caption{Mean, median and dispersion values for the best-fit parameters
$r_{\rm c}$ and $\beta$ estimated for objects in the chosen redshift 
bins.
}
\begin{tabular}{l c c }
 \hline \\ 
 range $z$ & $r_{\rm c}$ & $\beta$  \\ 
  & $h_{70}^{-1}$ kpc &  \\ 
  & &  \\ 
 \hline \\ 
 $0.4-0.6$ & 98, 137 (162) & 0.615, 0.635 (0.217) \\
 $0.6-0.8$ & 111, 138 (76) & 0.639, 0.627 (0.096) \\
 $0.8-1.0$ & 121, 166 (166) & 0.542, 0.682 (0.324) \\
 $> 1.0$ & 137, 128 (61) & 0.808, 0.778 (0.333) \\
 \hline \\ 
\end{tabular}

\label{tab:rcb}
\end{table}

The spectra are extracted out to a maximum radius $R_{\rm spec}$,
which is chosen cluster by cluster so as to optimize the
signal-to-noise ratio in the 0.5--5 keV band where most of the \chandra
effective area is available. 
They are modeled between 0.6 and 8 keV with an
absorbed optically--thin plasma ({\tt tbabs(mekal)} in XSPEC v.~11.2,
Arnaud 1996) with redshift fixed to the value available in literature
and absorption equal to the Galactic value (as recovered from radio HI
maps, Dickey \& Lockman 1990) in correspondence of the X-ray peak.  A
local background obtained from regions of the same CCD free of any
point source is used.  The gas temperature, $T_{\rm gas}$, metallicity
and the normalization $K$ of the thermal component are the only free
parameters of the spectral fit performed with a Cash statistics due to
the low number of counts per bin available (e.g. Nousek \& Shue 1989;
see also discussion on background-subtracted spectra at 
{\tt http://heasarc.gsfc.nasa.gov/ docs/ xanadu/ xspec/ manual/ node57.html}).

The surface brightness profiles are obtained from the
exposure-corrected image by fixing the number of counts per bin between
50 and 200, depending on the photons available, and is then fitted with
an isothermal $\beta-$model (Cavaliere \& Fusco-Femiano 1976).  This
model provides a reasonably good fit to all the profiles (see plots in
Appendix), apart from seven clusters 
whose measured profiles have a probability $<0.01$ not to be a random
realization of a $\beta$--model, as drawn from a $\chi^2$ distribution
(RXJ1416, $\chi^2_{\rm red}$ = 1.98; RXJ1701 , $\chi^2_{\rm red}$ =
1.96; RXJ1525, $\chi^2_{\rm red}$ = 1.47; MS0451, $\chi^2_{\rm red}$ =
1.45; MS0016, $\chi^2_{\rm red}$ = 1.40; RXJ1113, $\chi^2_{\rm red}$ =
1.49; RXJ2302, $\chi^2_{\rm red}$ = 1.59).
This is not unexpected, being now well known that highly spatially
resolved observations of nearby systems have surface brightness profiles
poorly described by simple functional forms, due to the presence
of breaks and general irregularities that cannot be reproduced by any
smoothed curve (see, e.g., the case of A2390 in Allen, Ettori \& Fabian 2001
and the detected front in RXJ1252, Rosati et al. 2003) and overall
two-dimensional shape not strictly circular as here assumed 
(e.g. Maughan et al. 2003 on RXJ1113 and Donahue et al. 2003 on MS0451).
On the other hand, the results obtained from the application 
of the $\beta-$model represent a robust estimates of the effective 
values under exam, such as gas and total gravitating mass, as shown
for simulated clusters (e.g. Mohr, Mathiesen \& Evrard 1999).  

We have also tested a double $\beta-$model in the form suggested by
Ettori (2000 and reference therein), which provides a consistent 
physical framework for the modeling: 
$S_{\rm b} = S_{\rm 0, inn} (1-x^2)^{0.5+3\beta_{\rm inn}}
+S_{\rm 0} (1+x^2)^{0.5-3\beta} = r_{\rm c} \Lambda(T) n_{\rm 0,
gas}^2 [ f_{\rm inn} B(0.5, 3\beta_{\rm inn}+1)
(1-x^2)^{0.5+3\beta_{\rm inn}} \ + \ (1-f_{\rm inn}) B(3\beta-0.5,0.5)
(1+x^2)^{0.5-3\beta}]$, where $r_{\rm c}$ is the common core radius,
$x = r/r_{\rm c}$, $\Lambda(T)$ is the cooling function, $n_{\rm 0,
gas}$ is the central gas density and $B()$ is the incomplete beta
function.  From a statistical point of view, none of the clusters
requires to add two parameters (i.e. the slope $\beta_{\rm inn}$ of
the inner, corrected $\beta-$model and the relative contribution
$f_{\rm inn}$ of the central, cool component).  The F-test (Bevington
\& Robinson 1992, p.205ff) gives probability values in the range
between 0.40 (MS1054) and 0.85 (MS0016), so that we can conclude
that a double $\beta-$model is not required by our data at the fixed
threshold of 0.95 (95 per cent confidence level). 
In particular, we note
that RXJ1416 and RXJ1701, the two clusters with worst $\chi^2$ associated 
to the single $\beta-$model, show the smaller core radii, but still cannot 
accommodate a double $\beta-$model that provides a F-test statistic with 
probability of 0.82 and 0.67, respectively.  
On the other hand, few clusters at redshift around 0.3 not included in the 
present sample, such as MS1008 and MS2137, require a double $\beta-$model
with probability of $>$0.99 and 0.98, respectively.
It is worth noticing two items concerning the surface brightness 
of high--$z$ clusters. First, thanks to the \chandra spatial resolution,
we are in the condition to resolve on $\sim$ 20 kpc scale their inner regions.
For example, we are able to define between 4 and 7 spatial bins with a 
fixed number of 50 counts within 100 kpc in the four $z>1$ objects.
Then, the evidence that their surface brightness cannot accommodate
a double $\beta-$model indicates that these clusters do not show any  
significant excess in their central brightness, consistent with the
picture that systems at high redshift appear as structures in
formation and not completely relaxed.
We refer to a forthcoming paper (Ettori et al. in prep.) for
a more detailed analysis of the distribution of the observed surface 
brightness, in particular in the central regions, as function of 
the object's redshift.

We have investigated the presence of trends between the best-fit
parameters of the $\beta-$model, the redshift and the gas temperature
of the objects in the overall sample by evaluating the significance of 
the Spearman's rank correlation.  
We obtain a relevant deviation (Spearman's rank correlation with a
probability $P = 4 \times 10^{-5}$) from the no-correlation case only for
the well-known degeneracy between $r_{\rm c}$ and $\beta$ values,
being higher $\beta$ measured in correspondence of larger core radii
(see Fig.~\ref{fig:bmod_par}; for the same correlation observed in
nearby and intermediate-redshift objects see, e.g., Ettori \& Fabian 1999). 
A mild trend ($P = 0.03$) between $r_{\rm c}$ and $T_{\rm gas}$ is also
observed. The latter correlation has a milder correspondence between 
$\beta$ and $T_{\rm gas}$ ($P=0.17$) and indicates that larger core radii 
are measured in more massive systems, with no sizable change in the outer 
slope of the density profile. 
As shown in Fig.~\ref{fig:bmod_par} and quantified in Table~\ref{tab:rcb}, 
the best-fit parameters $r_{\rm c}$ 
and $\beta$ do not show any significant trend with the redshift.  

The advantage of using a $\beta-$model to parameterize the observed
surface brightness is that gas density and total mass profiles can be
recovered analytically and expressed by simple formula:
\begin{eqnarray}
n_{\rm gas}(r) & = & n_{0, \rm gas} (1+x^2)^{-3 \beta/2} \nonumber \\
M_{\rm tot}(<r) & = & \frac{3 \ \beta \ T_{\rm gas} \ r_{\rm c}}{G \mu m_{\rm p}}
 \frac{x^3}{ 1+x^2 },
\label{eq:beta}
\end{eqnarray}
where 
$x=r/r_{\rm c}$,
$\mu$ is the mean molecular weight in atomic mass unit ($=$0.6),
$G$ is the gravitational constant and $m_{\rm p}$ is the proton mass.

In particular, the central electron density is obtained from the
combination of the best-fit results from the spectral and imaging
analyses as follows:
\begin{equation}
n^2_{0, \rm ele} = \frac{4 \pi d_{\rm ang}^2 \times (1+z)^2 \times K 
 \times 10^{14}} {0.82 \times 4 \pi r_{\rm c}^3 \times EI},
\label{eq:ei}
\end{equation}
where $d_{\rm ang}$ is the angular--size distance,  $K = 10^{-14} \times 
\int n_{\rm p} n_{\rm e} dV / [4 \pi d_{\rm ang}^2 (1+z)^2 ]$ is the 
normalization of the thermal spectrum measured within XSPEC
and the emission integral $EI$ is estimated by integrating the emission 
from the spherical source out to 10 Mpc along the line of sight, $EI =
\int_0^{x_1} (1+x^2)^{-3\beta} x^2 dx + \int_{x_1}^{x_2}
(1+x^2)^{-3\beta} x^2 (1-\cos \theta) dx$, with $\theta =
\arcsin(R_{\rm spec}/r)$, $x_1 = R_{\rm spec}/r_{\rm c}$ and $x_2 = 10
{\rm Mpc} /r_{\rm c}$, $(\beta, r_{\rm c})$ are the best-fit
parameters of the $\beta-$model and we assume $n_{\rm p} = 0.82 n_{\rm
e}$ in the ionized intra-cluster plasma.  
To evaluate the relative errors on the estimated quantities, we
repeat the measurements after 1\,000 random selection of a temperature, 
normalization $K$ and surface brightness profile, which were drawn from Gaussian 
distributions with mean and variance in accordance to the best-fit results.
By this process, we obtain a distribution of the estimates of gas density,
gas mass and total mass. The median value and the 16th and 84th percentile of 
each distribution are adopted as central and $(-1, 1) \sigma$ interval,
respectively (see Table~\ref{tab:obj_a}).

\begin{figure}
\epsfig{figure=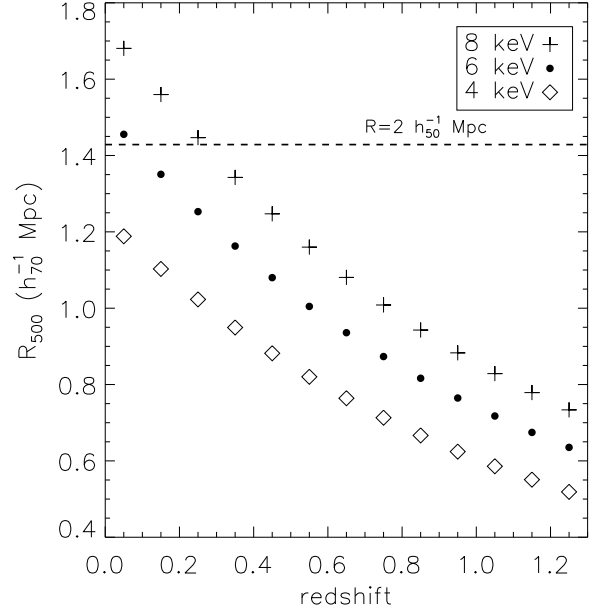,width=0.5\textwidth}
\caption{Values of $R_{500}$ as function of redshift for the cosmological
model assumed.
The conversion from temperature (mass) to $R_{\Delta}$ is obtained by
using the best-fit results of the $M-T$ relation (see Sect.3) and
equation~\ref{eq:delta}.
As comparison of what is adopted in other similar work (e.g. Vikhlinin et al. 
2002), we draw the value corresponding to a fixed radius of 2 $h_{50}^{-1}$ Mpc
(dashed line).
} \label{fig:rd} \end{figure}

\begin{figure}
\epsfig{figure=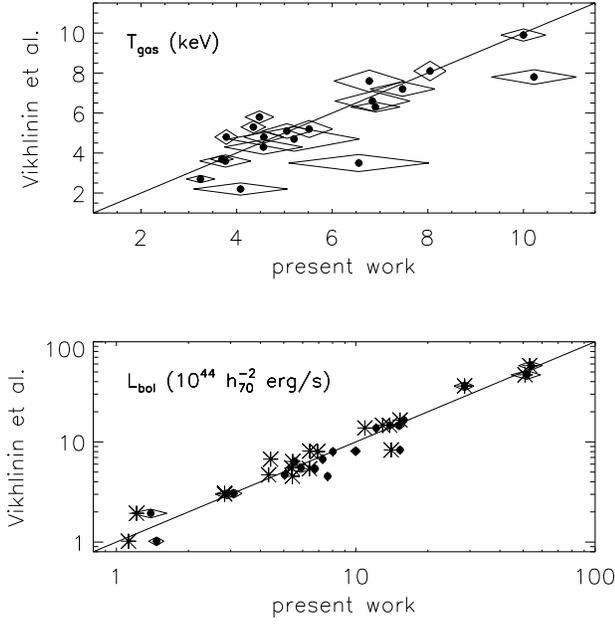,width=0.5\textwidth}
\caption{Comparison between the temperatures and luminosities estimated
in the present work and by Vikhlinin et al. (2002) 
for the 21 clusters that we have in common.  
({\it Upper panel}) A median (mean) difference of 4 (8) per cent is present 
between our measurements of gas temperature and the values quoted in 
 Vikhlinin et al.
({\it Lower panel}) The dots indicate the estimates within  $2 h_{50}^{-1}$ Mpc,
whereas the asterisks show our measurements of $L_{\rm bol} (<R_{500})$.
 A median (mean) difference of 0 (8) per cent is observed between our values
and the ones in Vikhlinin et al.
} \label{fig:v02}
\end{figure}

From the total mass profile, $M_{\rm tot}(<r)$, obtained for each
galaxy cluster observed at redshift $z$, we then evaluate $R_{\Delta}$
as the radius encompassing a
fixed density contrast, $\Delta_z$, with respect to the critical
density, $\rho_{\rm c, z}$, as
\begin{equation}
\Delta_z = \frac{3 M_{\rm tot}(<R_{\Delta})}{4\pi \rho_{\rm c, z}
  R^3_{\Delta}}.
\label{eq:delta}
\end{equation}

Observational results for nearby clusters (e.g. De Grandi \& Molendi
2002, Pratt \& Arnaud 2002) show that temperature profiles do not
decline significantly ($\la$ 20 per cent) out to an overdensity $\sim
500$ (for a universe with $\Omega_{\rm m}=$1 and
$\Omega_{\Lambda}=$0). Therefore, to make a proper use of our
single-temperature measurement, we estimate all the observed
quantities considered in this work, i.e.  total mass, gas mass and
luminosity to $R_{\Delta}=R_{500}$.  Specifically, an overdensity of
500 is assumed for an Einstein-de Sitter universe, whereas $\Delta_z =
500 \times \left[1 +82 \left(\Omega_z-1\right) / \left(18 \pi^2\right)
-39 \left(\Omega_z-1\right)^2 / \left(18 \pi^2\right) \right]$, with
$\Omega_z = \Omega_{\rm m} (1+z)^3/E_z^2$, is assumed accordingly to
the approximation of the solution for the collapse of a spherical
top-hat perturbation that just virialized at the observed redshift $z$
(e.g. Bryan \& Norman 1998).  Fig.~\ref{fig:rd} shows the dependence
of $R_{500}$ upon the typical mass (temperature) and redshift values
spanned by the clusters in our sample.  Changes of about 80
per cent are expected for $R_{500}$ in the redshift range 0.4 and 1.3,
whereas a fixed metric radius remains constant by definition.

In this work, the luminosity within $R_{\Delta}$ is then estimated by 
correcting the value measured directly from the spectrum within $R_{\rm spec}$
as follows
\begin{equation}
L(<R_{\Delta}) = L(<R_{\rm spec}) \ \frac{\int_0^{R_{\Delta}} 
(1+x^2)^{-3 \beta} x^2 dx}{EI},
\label{eq:lumx}
\end{equation}
where $EI$ is given in Equation~\ref{eq:ei}.  No further corrections
on the measurements of the luminosity are applied, i.e. no excess from
any centrally peaked surface brightness due to the presence of
``cooling flows'' is excised.  On the other hand, the lack of any
evidence for a significant excess of the brightness in the core
regions, as discussed above, indicates that no relevant differences in
the total luminosity are expected when excision is applied
(e.g. Vikhlinin et al. 2002) as we also show at the end of this
section.

The quoted errors on luminosity are obtained through the propagation
of the errors obtained in XSPEC from the distribution around the
best-fit values in the spectral analysis (which are the dominant
contribution), and the uncertainties related to the $\beta-$model in
moving from $R_{\rm spec}$ to $R_{\Delta}$.

In the Appendix, we show the plots of the surface brightness profiles
of our clusters, along with the $\beta-$model obtained through a
least-squares fitting process and the relative positions of
$R_{\Delta}$, $R_{\rm spec}$ and $R_{2 \sigma}$. The latter radius
indicates up to where the signal is above $2 \sigma$ with respect to
the best-fit value of the background and gives an indication of the
region of the cluster where a diffuse emission is detectable.

Out of 28 clusters, 17 have $R_{2 \sigma}$ that is smaller than
$R_{500}$ by a factor between 1.1 and 3.0 (median of 1.4) and need
extrapolation to compute their quantities. The spectral aperture,
$R_{\rm spec}$, corresponds roughly to an overdensity of $\sim$ 2000
and is about 0.6$\,R_{2 \sigma}$ and 3.6 times larger than the
cluster core radius (median values).  It is worth noticing that more
than 75 per cent of the cluster luminosity, on average, is however sampled 
within $R_{\rm spec}$ because of the $n_{\rm gas}^2$ dependence of the X-ray
thermal emission.

Finally, we compare our estimates of $T_{\rm gas}$ and $L_{\rm bol}$
with the values quoted by Vikhlinin et al. (2002). These authors have
21 clusters in common with us (including also CL0024 that is not
listed here because at redshift 0.395 and, therefore, below our cut at
$z=0.4$).  The agreement is remarkably good, with a median ratio
between our and Vikhlinin et al. measurements of temperature of
1.04 (mean: 1.08; see Fig.~\ref{fig:v02}) and estimates of
luminosity (at $2 h_{50}^{-1}$ Mpc) of 1.00 (1.08 in average). Larger
($> |2 \sigma|$ when comparing our values vs Vikhlinin et al. ones)
deviations are observed in the temperature estimates of RXJ1347 ($-4.7 \sigma$), 
RXJ1701 ($-3.0 \sigma$), CL0024 ($-2.3 \sigma$), 3C295 ($-2.3 \sigma$)
and MS1054 ($+2.5 \sigma$), and in the
luminosity measurements of 3C295 ($+8.2 \sigma$), RXJ1416 ($+6.2 \sigma$), 
RXJ1317 ($+3.1 \sigma$) and MS0302 ($+2.3 \sigma$). 
We notice that all these objects (apart from MS1054 that presents a significant 
substructure that we mask out in our analysis as described in the caption of Table~1)
are classified by Vikhlinin and collaborators as systems 
with a ``possible cooling core'' (even though they do not describe how
they quantify the sharpener of the central brightness). 
The core emission is then excised by Vikhlinin et al. during the process
of the determination of the gas temperature and luminosity, 
whereas we maintain it in our procedure.

\begin{figure*}
\hbox{
 \epsfig{figure=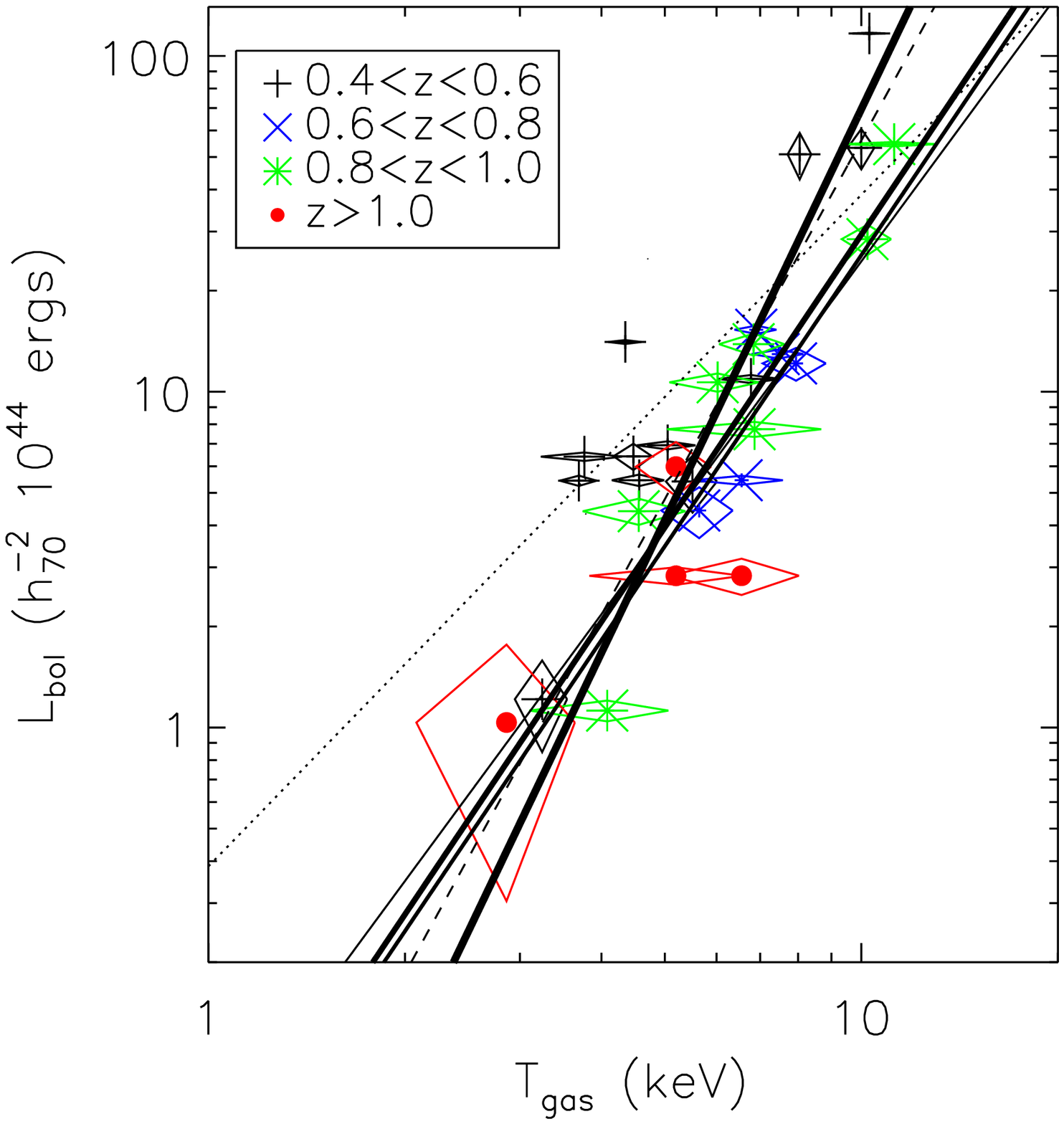,width=0.5\textwidth}
\epsfig{figure=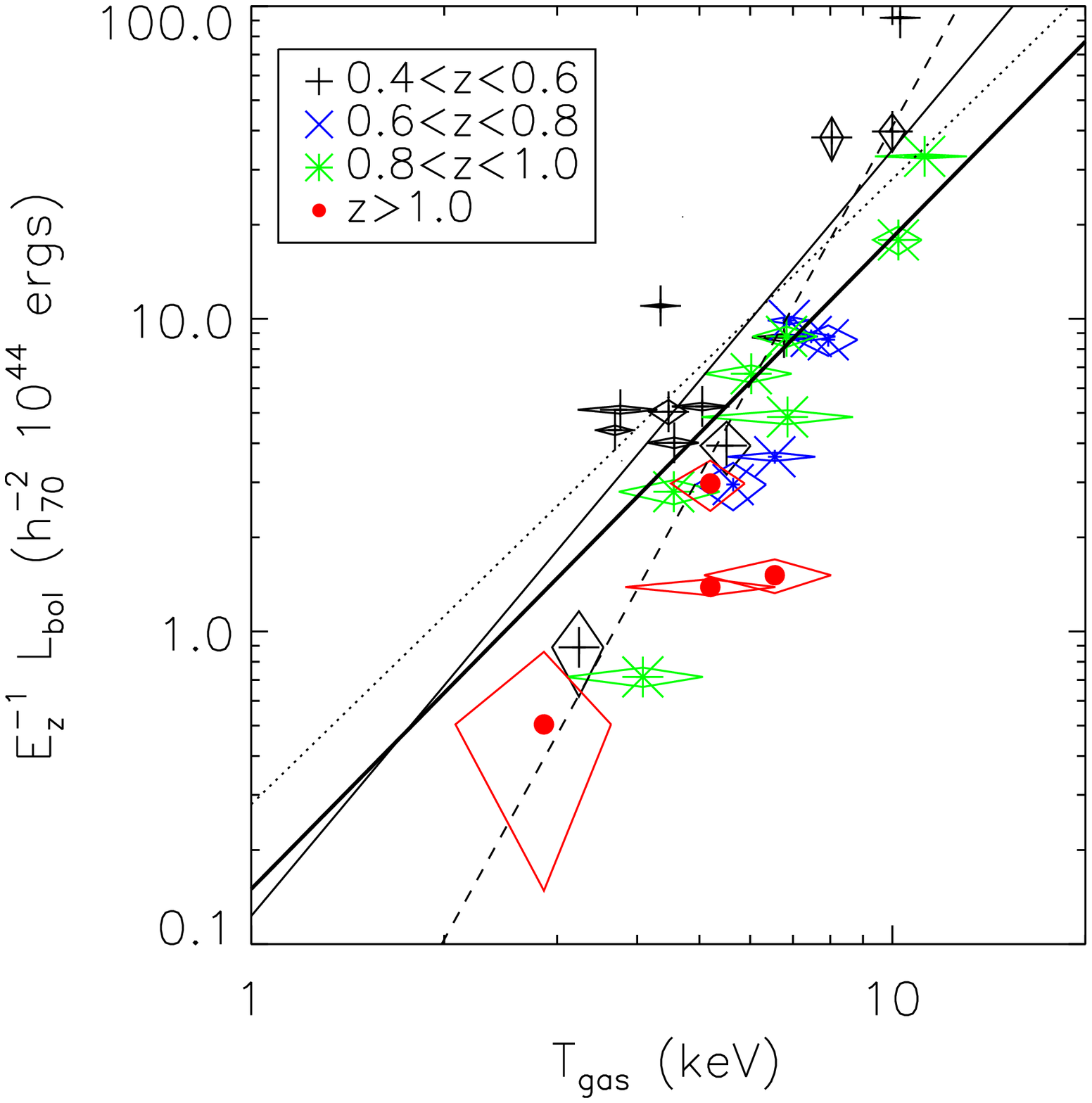,width=0.5\textwidth}
} \hbox{ 
 \epsfig{figure=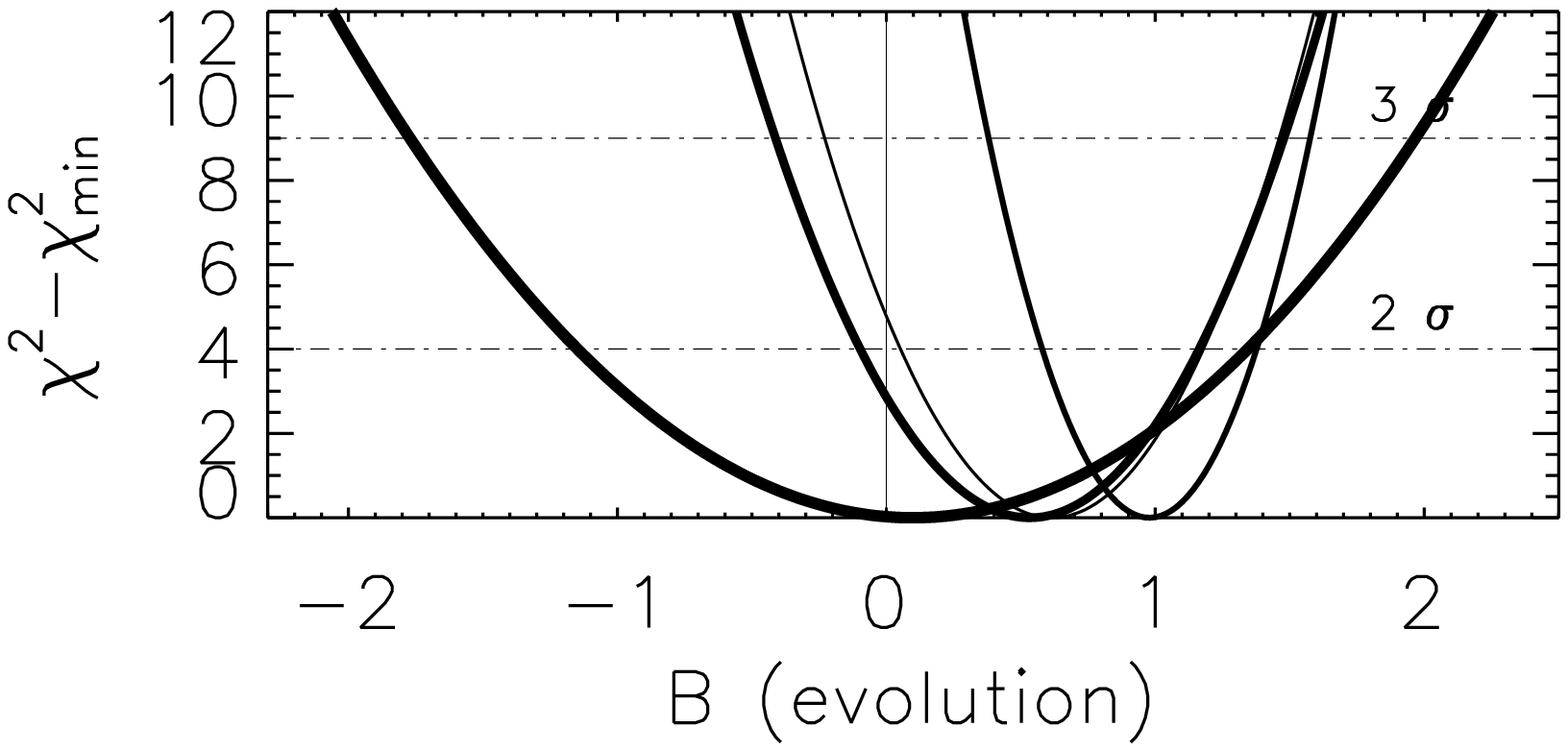,width=0.5\textwidth}
 \epsfig{figure=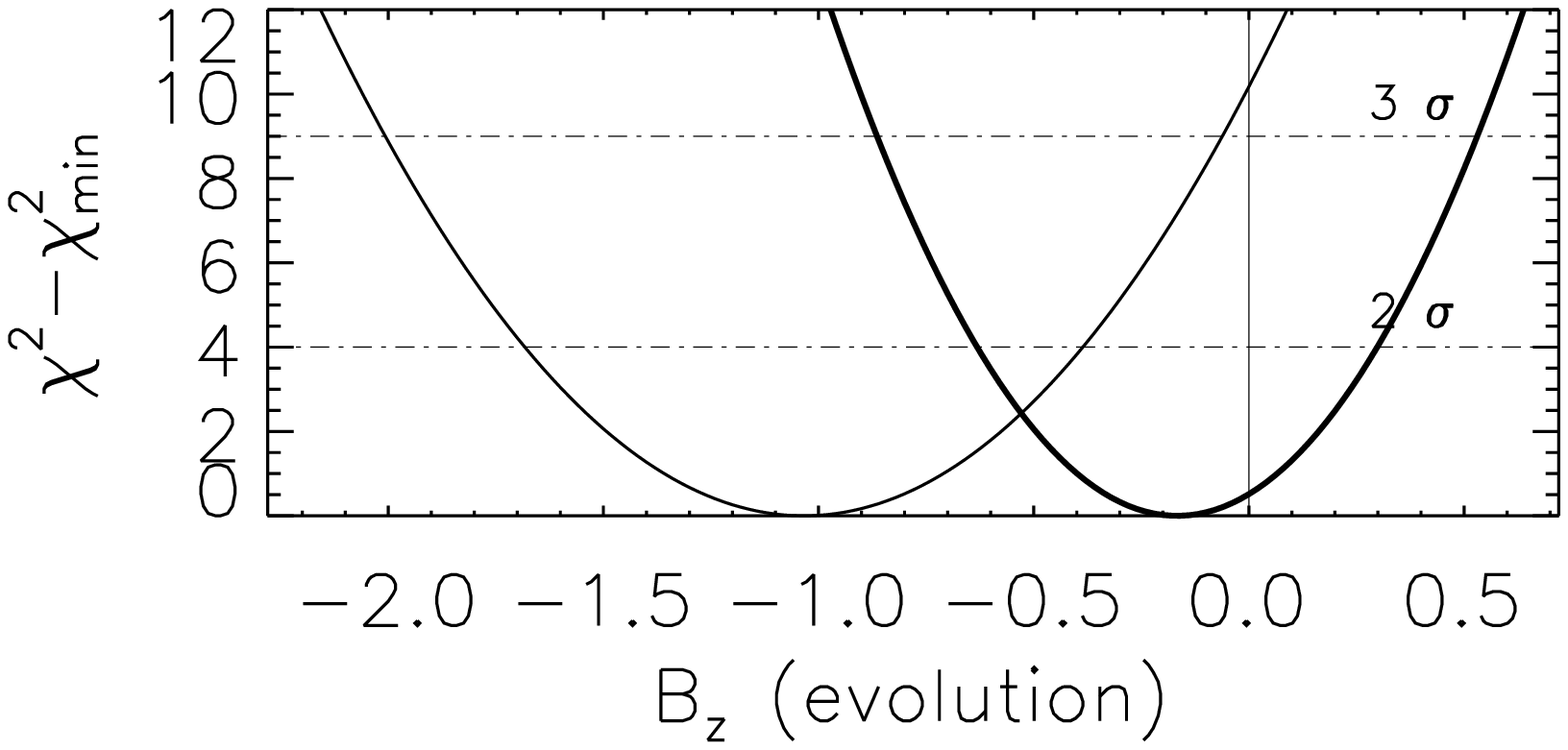,width=0.5\textwidth} }
\caption{$L-T$ relation without ({\it left}) and with ({\it right})
correction by $E_z$. 
{\it (Upper panels)} Dotted line: slope fixed to the predicted value of $2$.
Dashed line: slope free. The solid lines represent the local best-fit
results (from thinnest to thickest line): 
Markevitch (1998; with core excised in cooling flows clusters),
Arnaud \& Evrard (1999), Novicki et al. (2002; for objects at $z<0.3$)
and Novicki et al. (2002; for objects at $0.6 > z > 0.3$).
Solid line: Ettori et al. (2002; thinest line), Allen et al. (2001).
{\it (Bottom panels)} Plot of the $\Delta \chi^2$ distribution for the one
interesting parameter $B (B_z)$ given in eqn.~\ref{eq:chi2}.
Each solid line corresponds to a local scaling relation plotted with the same
thickness in the upper panel and compared to all the 28 objects with $z \ge 0.4$
in our sample.  The $2$ and $3 \sigma$ limits are shown as dot-dashed lines. 
} \label{fig:lt} \end{figure*}


\section{The scaling relations at $0.4 < z < 1.3$}

\begin{table}
\caption{Best-fit results on the scaling relations corrected by the cosmological
factor $E_z$
(dashed and dotted lines in the plots; see equation~\ref{eq:fit}).
The temperature, $T_{\rm gas, 6}$, is in unit of 6 keV;
the luminosity, $L_{\rm bol, 44}$, in $10^{44} h_{70}^{-2}$ erg s$^{-1}$;
the total mass, $M_{\rm tot, 14}$, in $10^{14} h_{70}^{-1} M_{\odot}$;
the gas mass, $M_{\rm gas, 13}$, in $10^{13} h_{70}^{-5/2} M_{\odot}$.
When the slope $A$ is fixed, we estimate the error-weighted mean
of $(\log Y -A \log X)$ and evaluate the error after resampling $Y$ and $X$
by 1\,000 times according to their uncertainties.
The scatter on $Y$ is measured as $\left[ \sum_{j=1,N}   
\left(\log Y_j -\alpha -A \log X_j \right)^2 /N \right]^{1/2}$.
Note that the scatter along the X-axis can be estimated as
$\sigma_{\log X} = \sigma_{\log Y} / A$.
} \begin{tabular}{l@{\hspace{.7em}} c@{\hspace{.7em}} c@{\hspace{.7em}} c@{\hspace{.7em}} }
 \hline \\ 
 relation $(Y-X)$ & $\alpha$ & $A$ & $\sigma_Y$ \\ 
  & &  & \\ 
 \hline \\ 
 $E_z^{-1} L_{\rm bol, 44}-T_{\rm gas, 6}$ & $0.79 (\pm0.07)$ & $3.72 (\pm0.47)$ & $0.35$ \\ 
 & $1.00 (\pm0.02)$ & $2.00$ (fixed) & $0.41$ \\ 
  & &  & \\ 
 $E_z M_{\rm tot, 14}-T_{\rm gas, 6}$ & $0.75 (\pm0.03)$ & $1.98 (\pm0.30)$ & $0.15$ \\ 
 & $0.74 (\pm0.02)$ & $1.50$ (fixed) & $0.15$ \\ 
  & &  & \\ 
 $E_z M_{\rm gas, 13}-T_{\rm gas, 6}$ & $0.79 (\pm0.03)$ & $2.37 (\pm0.24)$ & $0.17$ \\ 
 & $0.92 (\pm0.01)$ & $1.50$ (fixed) & $0.22$ \\ 
  & &  & \\ 
 $E_z^{-1} L_{\rm bol, 44}- E_z M_{\rm tot, 14}$ &$-0.63 (\pm0.32)$ & $1.88 (\pm0.42)$ & $0.47$ \\ 
 & $0.09 (\pm0.03)$ & $1.33$ (fixed) & $0.52$ \\ 
 \hline \\ 
\end{tabular}

\label{tab:fit}
\end{table}

A sensible way to test the hierarchical formation of galaxy clusters
and the influence of non-gravitational processes on their
self-similarity at different epochs is to study how the scaling
relations behave and evolve at high redshifts. 
In practice, by using the observed ($L, T$) and derived ($M_{\rm tot}, 
M_{\rm gas}$) physical properties measured within a given overdensity 
$\Delta_z$ with respect to the critical overdensity $\rho_{\rm c,z}$,
one wants to investigate whether the quantities $E_z M_{\rm tot} /
T_{\rm gas}^{3/2}$ and $E_z^{-1} L_{\rm bol} / T_{\rm gas}^2$ [and the
complementary ones, $E_z M_{\rm gas} / T_{\rm gas}^{3/2}$ and
$E_z^{-1} L_{\rm bol} / (E_z M_{\rm tot})^{4/3}$] show, or not, 
any evolution with redshift and/or dependence upon one of the physical
properties, typically the temperature.

\begin{figure}
\vbox{
\epsfig{figure=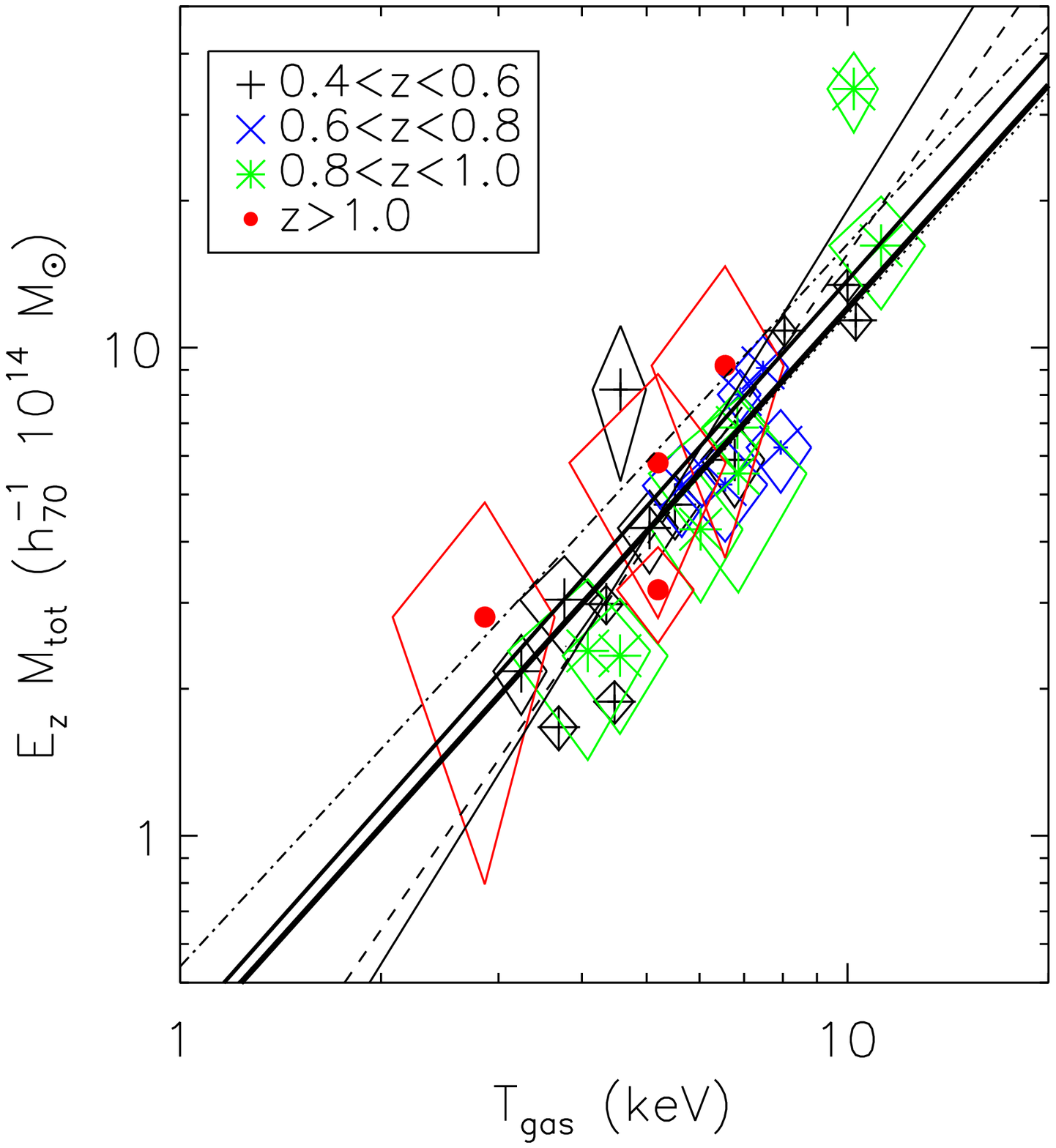,width=0.5\textwidth}
\epsfig{figure=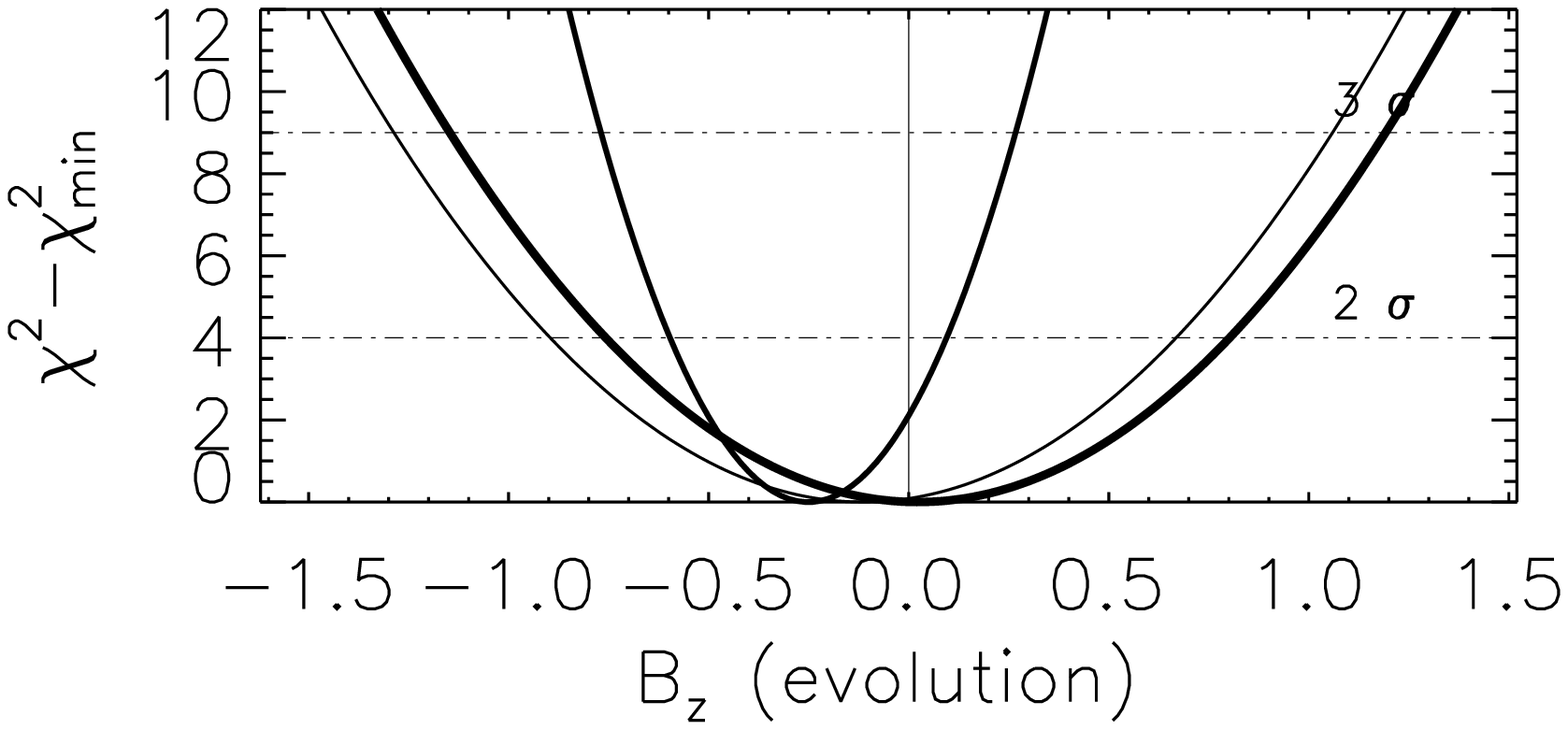,width=0.5\textwidth} }
\caption{$M_{\rm tot}-T$ relation. 
{\it (Upper panels)} Dotted line: slope fixed to the predicted value of $3/2$.
Dashed line: slope free. The solid lines represent the local best-fit
results (from thinnest to thickest line): Ettori et al. (2002),
Finoguenov et al. (2001), Allen et al. (2001).
The dash-dot line indicates the best-fit results from hydrodynamical
simulations in Mathiesen \& Evrard (2001, Table~1, with temperatures
from simulated spectral analysis in the band 0.5--9.5 keV).
{\it (Bottom panels)} Plot of the $\Delta \chi^2$ distribution for the one
interesting parameter $B_z$ given in eqn.~\ref{eq:chi2}.
Each solid line corresponds to a local scaling relation plotted with the same
thickness in the upper panel and compared to all the 28 objects with $z \ge 0.4$
in our sample.  The $2$ and $3 \sigma$ limits are shown as dot-dashed lines.
} \label{fig:mt} \end{figure}

\begin{figure}
\vbox{
  \epsfig{figure=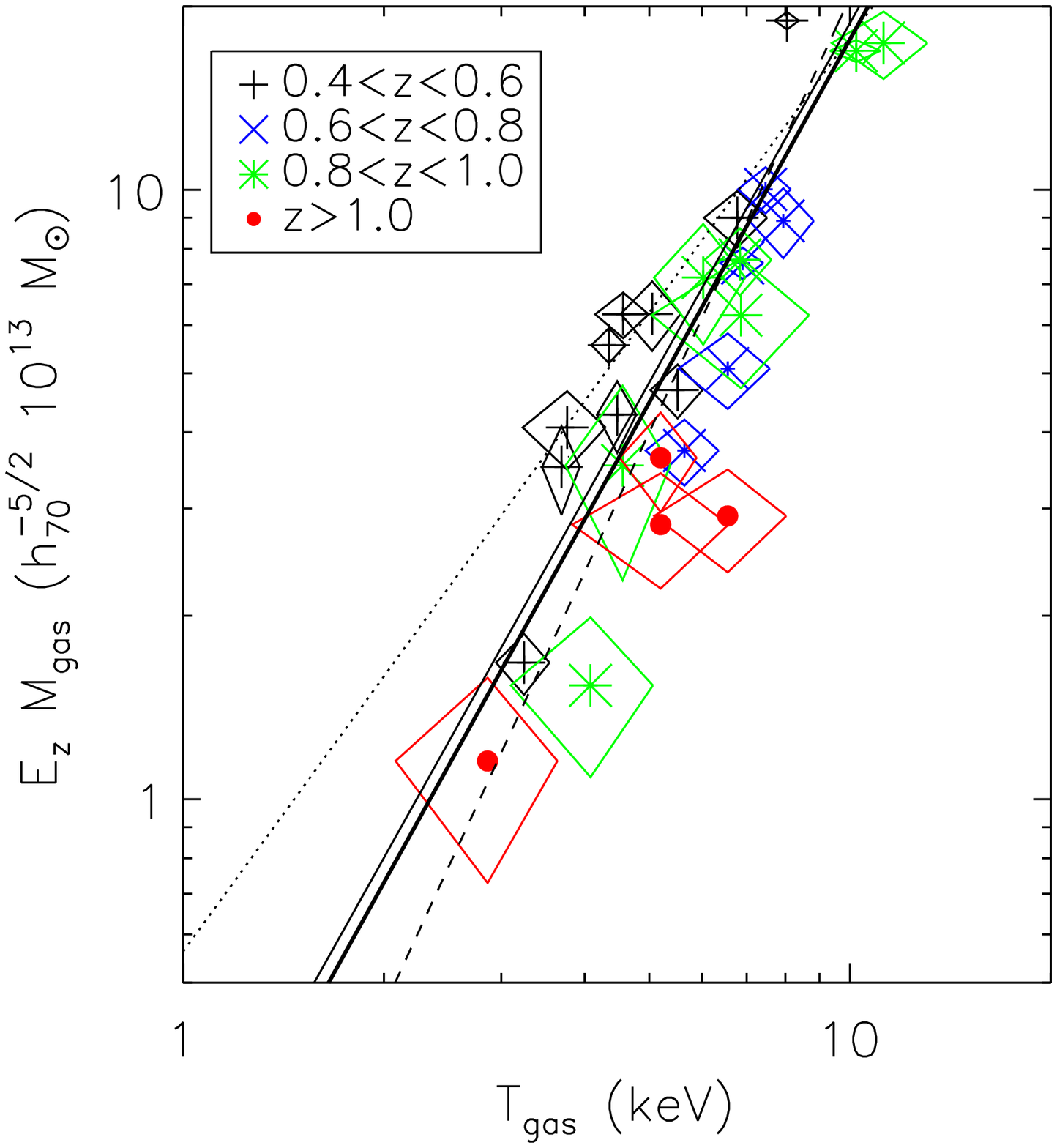,width=0.5\textwidth}
  \epsfig{figure=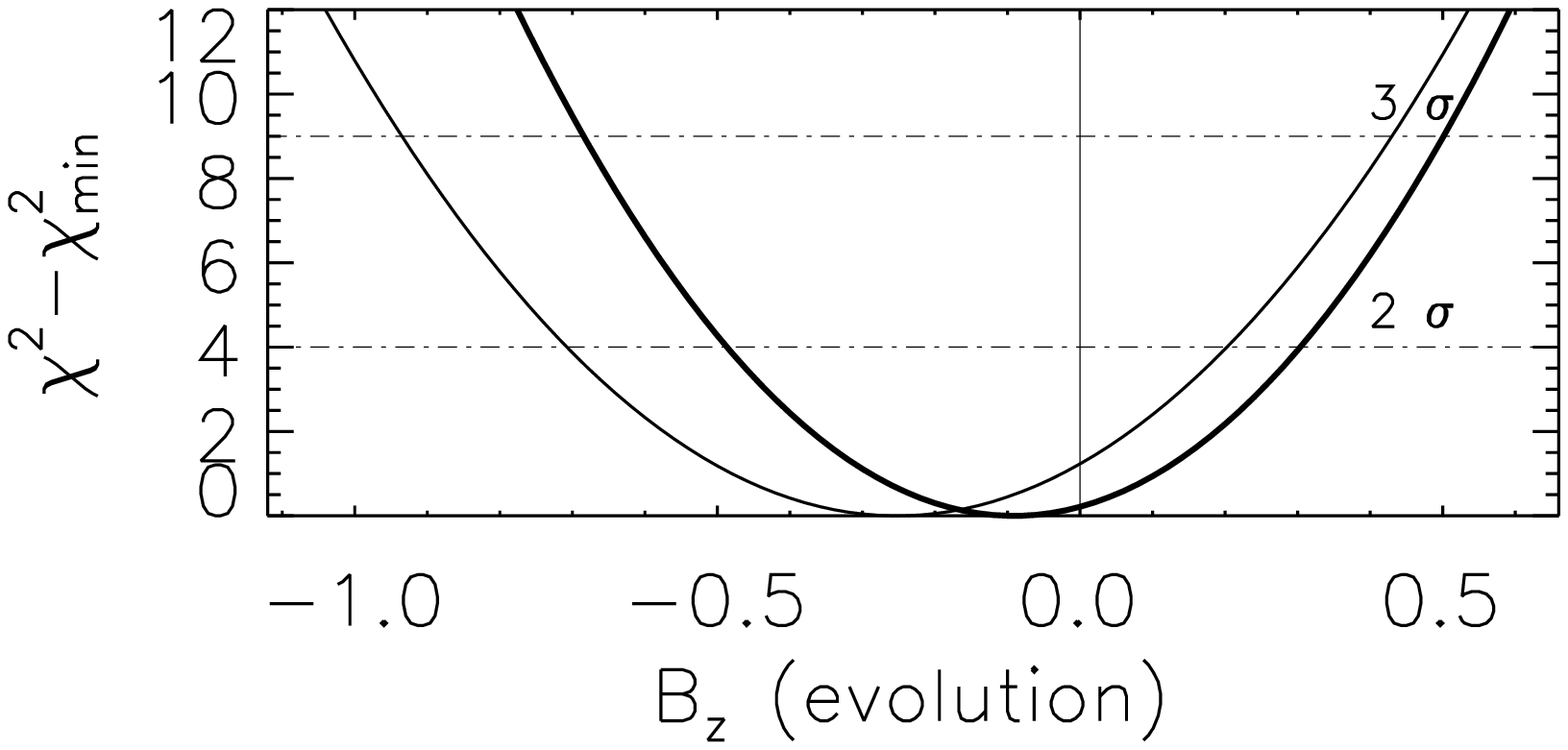,width=0.5\textwidth}
} \caption{ $M_{\rm gas}-T$ relation. 
{\it (Upper panels)} Dotted line: slope fixed to the predicted value of $3/2$.
Dashed line: slope free. The solid lines represent the local best-fit
results (from thinnest to thickest line): 
Ettori et al. (2002), Mohr et al. (1999).
{\it (Bottom panels)} Plot of the $\Delta \chi^2$ distribution for the one
interesting parameter $B_z$ given in eqn.~\ref{eq:chi2}.
Each solid line corresponds to a local scaling relation plotted with the same
thickness in the upper panel and compared to all the 28 objects with $z \ge 0.4$
in our sample.  The $2$ and $3 \sigma$ limits are shown as dot-dashed lines.
} \label{fig:mgt} \end{figure}

\begin{figure}
\vbox{
  \epsfig{figure=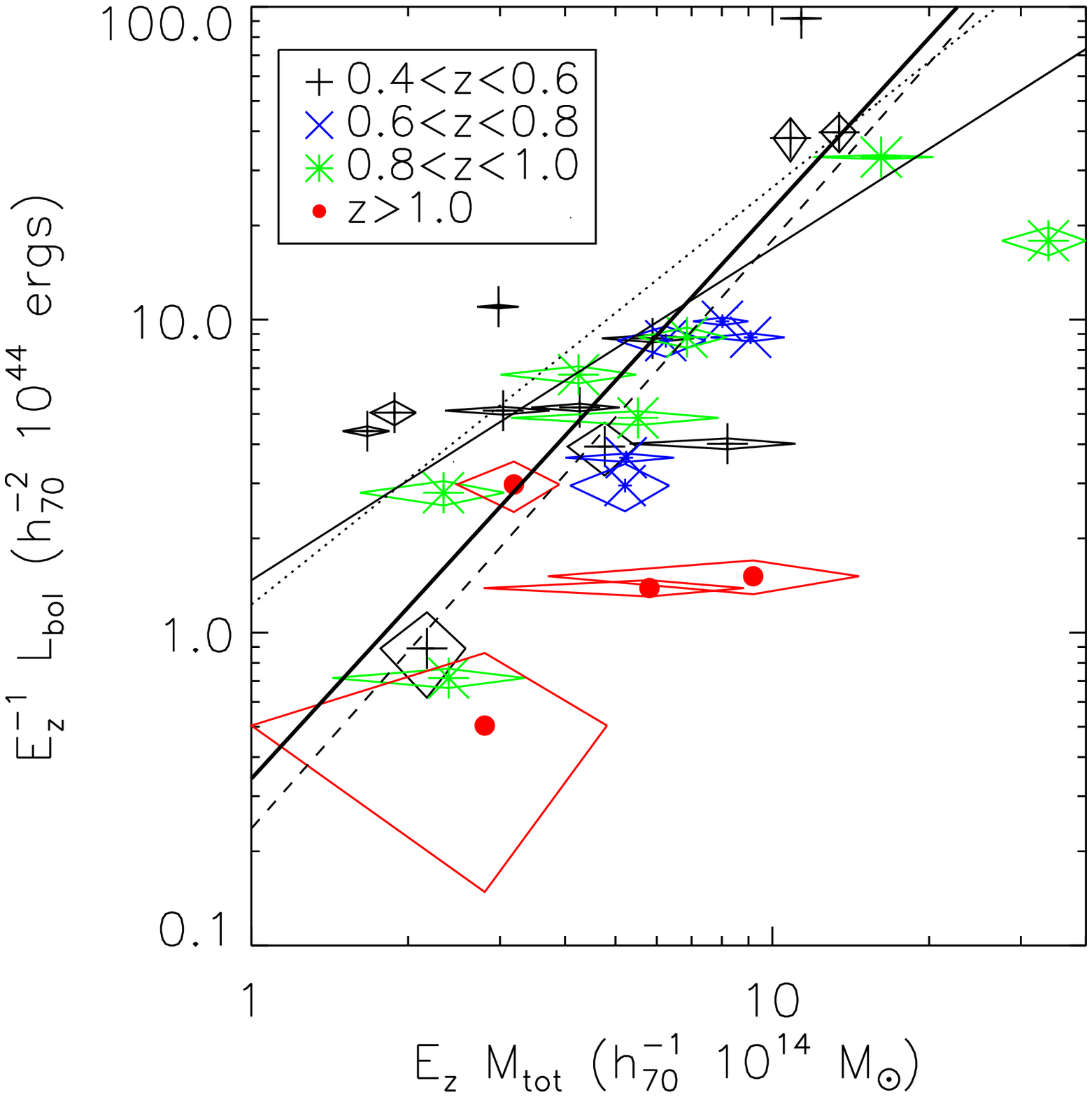,width=0.5\textwidth}
  \epsfig{figure=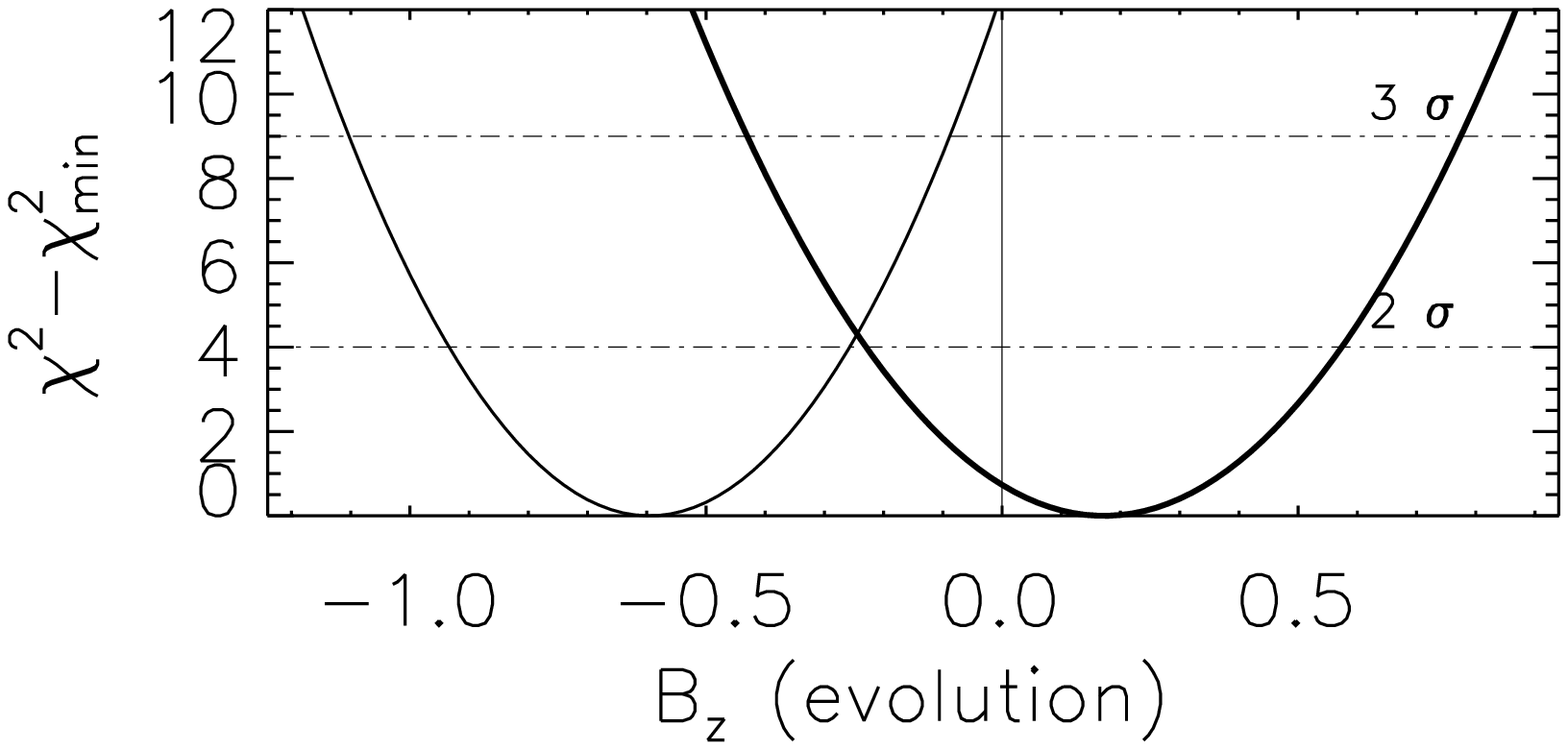,width=0.5\textwidth}
} \caption{ $L-M_{\rm tot}$ relation.
{\it (Upper panels)} Dotted line: slope fixed to the predicted value of $4/3$.
Dashed line: slope free. The solid lines represent the local best-fit
results (from thinnest to thickest line):
Ettori et al. (2002), Reiprich \& B\"ohringer (2002).
{\it (Bottom panels)} Plot of the $\Delta \chi^2$ distribution for the one
interesting parameter $B_z$ given in eqn.~\ref{eq:chi2}.
Each solid line corresponds to a local scaling relation plotted with the same
thickness in the upper panel and compared to all the 28 objects with $z \ge 0.4$
in our sample.  The $2$ and $3 \sigma$ limits are shown as dot-dashed lines.
} \label{fig:lm} \end{figure}

The behavior of these scaling laws are examined first 
in their normalization and slope by fitting the logarithmic relation
\begin{equation}
\log Y = \alpha +A \log X
\label{eq:fit}
\end{equation}
between two sets of measured quantities $\{X_j\}$ and $\{Y_j\}$.
We use the bisector modification (i.e. the best-fit results bisect 
those obtained from minimization in vertical and horizontal directions) 
of the linear regression algorithm in Akritas \& Bershady (1996 and 
references therein, hereafter BCES) that takes into account both any 
intrinsic scatter and errors on the two variables considered as symmetric.
The uncertainties on the best-fit results are obtained
from 10\,000 bootstrap resampling.
The results on the best-fit normalization and slope for the scaling
laws here investigated are quoted in Table~\ref{tab:fit} and shown
as dotted (when the slope is fixed to the predicted value) and dashed 
(when the slope is a free parameter) lines in Figures~~\ref{fig:lt},
\ref{fig:mt}, \ref{fig:mgt} and \ref{fig:lm}.

\begin{figure}[h]
\epsfig{figure=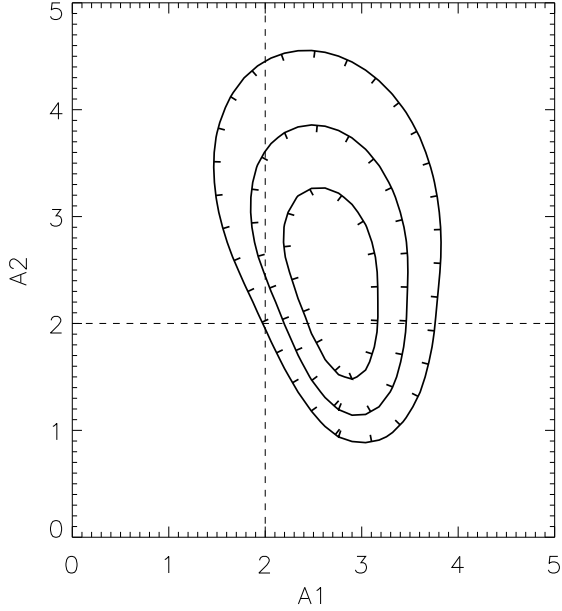,width=0.5\textwidth}
\caption{Slopes in the relation
$L_{\rm bol, 44} = A0 \times T_{\rm gas, 6}^{A1} (M_{\rm gas, 13}
/M_{\rm tot, 14})^{A2} (1+z)^{A3}$ obtained by minimization of the $\chi^2$
(best-fit results: $A0=2.0$, $A1=2.8$, $A2=1.9$, $A3=0.9$,
$\chi^2_{\rm min} =29.2$ with 24 d.o.f.) and after marginalization
on the parameters $A0$ and $A3$. The dashed lines indicate the predicted
values of 2 from the simple gravitational collapse model assuming
bremsstrahlung emission.
} \label{fig:ltm} \end{figure}

In general, we observe a steeper slope than expected from self-similar
model in all the scaling laws under exam, with $L-T$
(Fig.~\ref{fig:lt}) and $M_{\rm gas}-T$ (Fig.~\ref{fig:mgt}) relations
being the ones with deviations larger than $3 \sigma$ (best-fit of $A=
3.72\pm0.47$ and $2.37\pm0.24$, respectively), whereas $M_{\rm tot}-T$
(Fig.~\ref{fig:mt}) and $L-M_{\rm tot}$ (Fig.~\ref{fig:lm}) deviates
by about $1.5 \sigma$ from the expected values of $A=1.5$ and
$1.33$. This is a trend usually well-observed at lower redshift and
that affects all the quantities depending directly on the measurements
of the amount of gas, like luminosity and gas mass (e.g. Edge \&
Stewart 1991, Fabian et al. 1994, Mushotzky \& Scharf 1997, Allen \&
Fabian 1998, Mohr et al. 1999, Novicki et al. 2001, McCarthy et
al. 2002, Ettori et al. 2002).  This confirms that simple
gravitational collapse is not the only process that governs the
accretion of baryons in the potential well, but some combination of
properly efficient cooling and extra heating is required to take place
before or during the hierarchical formation of clusters (Kaiser 1991, 
Evrard \& Henry 1991, Cavaliere, Menci \& Tozzi 1999, Tozzi \& Norman 2001,
Balogh et al. 2002, Voit et al. 2002, Tornatore et al. 2003).  Scaling
the X-ray luminosity as $L \propto n_{\rm gas}^2 T^2 R^3 \sim 
f_{\rm gas}^2 T^2$, David et al. (1993) suggested that if the gas mass 
fraction, $f_{\rm gas}=M_{\rm gas}/M_{\rm tot}$, is proportional 
to $T^{0.5}$,
the steepening of the $L-T$ relation can be explained.  However, we
rule out any statistically significant dependence of $f_{\rm gas}$ on
the gas temperature in two ways. First, we measure a
probability of 66 per cent that no-correlation is present in the
distribution of the values in the plane $f_{\rm gas}-T$ once the
Spearman's rank coefficient is estimated (by fitting a power-law to
the data, we measure a slope of $0.5 \pm 1.0$). Then, we try to
constrain the best-fit parameters of the model $L \sim T^{A1} (M_{\rm
gas}/M_{\rm tot})^{A2}$ given the estimated gas luminosities,
gas masses and total gravitating masses. As
shown in Fig.~\ref{fig:ltm}, $A1$ is consistent with a value of 3 and
different from 2 by $\sim 3 \sigma$, whereas $A2$ is in agreement with
the predicted value of 2 within $1 \sigma$ ($A2 = 2.48^{+0.42}_{-0.70}$).

As originally pointed out by Horner et al. (1999) and Nevalainen et
al. (2000), samples of nearby clusters indicate that systems at given
temperature are systematically less massive than the counterparts
measured in numerical simulations (e.g. Evrard et al. 1996, Mathiesen
\& Evrard 2001, Borgani et al. 2002), where normalizations of the
$M_{\rm tot}-T$ relation are larger by about 30 per cent or more than
observed, once the slope is fixed to the expected value of 3/2.
We confirm this result at $z>0.4$, where we measure a normalization of
$(0.37 \pm 0.02) \times 10^{14} h_{70}^{-1} M_{\odot}$ keV$^{-3/2}$,
that is indeed $\sim$ 30 per cent lower than what obtained in the
hydrodynamical simulations of Mathiesen \& Evrard (2001; see
Fig.~\ref{fig:mt}), but well in agreement with previous observational
results obtained for cluster samples at lower redshifts
(e.g. Nevalainen et al. 2000, Allen, Schmidt \& Fabian 2001).

We note that the observed scatter in these relations indicates a
correlation between total/gas mass and gas temperature, which is as
tight at high redshift as measured locally. This suggests that a
marginal role is played by evolutionary effects, such as merging
history and non-gravitational heating, during the assembly of
baryonic and dark matter into the clusters.  A larger scatter is
present in the $L-T$ and $L-M_{\rm tot}$ relations. This is just the
consequence of the higher sensitivity of X--ray luminosity to the
details of the gas distribution and how it is affected by both
gravitational dynamics and non--gravitational processes.

\section{On the evolution of the scaling laws}

\begin{table*}
\begin{center}
\caption{
The best-fit values of the evolution parameter $B, B_z$ (see equations~\ref{eq:bfit} 
and \ref{eq:chi2} and comments in Section~4) are obtained by comparing to the 
local best-fit results, indicated in the second column and discussed in the text, 
the 28, 16 and 11 clusters with redshift larger than
$0.4$, $0.6$ and $0.8$, respectively.
The case of {\it no evolution} corresponds to a value of $B_z$ equal to 0.
The $L-T$ relation has been studied also not including the correction by $E_z$.
This relation is then expected to be proportional to $E_z$ and $B$ should
be measured in the range 0.6--0.9 in the adopted cosmology.
} \begin{tabular}{l@{\hspace{.7em}} c@{\hspace{.7em}} c@{\hspace{.7em}} c@{\hspace{.7em}} 
  c@{\hspace{.7em}} c@{\hspace{.7em}} }
 \hline \\ 
 relation & local ref. & & $(z\ge0.4)$ & $(z\ge0.6)$ & $(z\ge0.8)$\\
  & &  & \\ 
 \hline \\ 
 & & & \multicolumn{3}{c}{$B$} \\ 
  & &  & \\ 
 $L-T$ & Markevitch 98 & & $0.62 (\pm0.28)$ & $0.04 (\pm0.33)$ & $0.04 (\pm0.39)$ \\
 (fig.~\ref{fig:lt}) & Arnaud \& Evrard 99 & & $0.98 (\pm0.20)$ & $0.22 (\pm0.25)$ & $0.24 (\pm0.30)$ \\
 & Novicki et al. 02 ($z<0.3$) & & $0.54 (\pm0.31)$ & $-0.06 (\pm0.37)$ & $-0.02 (\pm0.43)$ \\
 & Novicki et al. 02 ($0.3<z<0.6$) & & $0.10 (\pm0.62)$ & $-0.48 (\pm0.73)$ & $-0.34 (\pm0.83)$ \\
  & &  & \\ 
  & &  & \\ 
 & & & \multicolumn{3}{c}{$B_z$} \\ 
  & &  & \\ 
 $E_z^{-1} L-T$ & Ettori et al. 02 & & $-1.04 (\pm0.32)$ & $-1.48 (\pm0.38)$ & $-1.46 (\pm0.44)$ \\
 (fig.~\ref{fig:lt}) & Allen et al. 01 & & $-0.16 (\pm0.23)$ & $-0.72 (\pm0.28)$ & $-0.78 (\pm0.32)$ \\
  & &  & \\ 
 $E_z M_{\rm tot}-T$ & Ettori et al. 02 & & $-0.12 (\pm0.39)$ & $-0.14 (\pm0.46)$ & $-0.06 (\pm0.52)$ \\
 (fig.~\ref{fig:mt}) & Finoguenov et al. 01 & & $-0.26 (\pm0.17)$ & $-0.16 (\pm0.21)$ & $-0.12 (\pm0.26)$ \\
 & Allen et al. 01 & & $0.02 (\pm0.39)$ & $0.04 (\pm0.45)$ & $0.06 (\pm0.52)$ \\
  & &  & \\ 
 $E_z M_{\rm gas}-T$ & Ettori et al. 02 & & $-0.26 (\pm0.23)$ & $-0.54 (\pm0.27)$ & $-0.54 (\pm0.32)$ \\
 (fig.~\ref{fig:mgt}) & Mohr et al. 99 & & $-0.10 (\pm0.20)$ & $-0.42 (\pm0.24)$ & $-0.42 (\pm0.28)$ \\
  & &  &\\ 
 $E_z^{-1} L-E_z M_{\rm tot}$ & Ettori et al. 02 & & $-0.60 (\pm0.17)$ & $-1.10 (\pm0.20)$ & $-1.12 (\pm0.25)$ \\
 (fig.~\ref{fig:lm}) & Reiprich \& B\"ohringer 02 & & $0.18 (\pm0.20)$ & $-0.86 (\pm0.25)$ & $-0.78 (\pm0.32)$ \\
  & &  & \\ 
 \hline \\ 
\end{tabular}

\end{center}
\label{tab:bfit}
\end{table*}

To constrain the evolution in the considered scaling laws, 
we fix $(\alpha, A)$ to the best-fit results obtained
from a sample of objects observed at lower redshift, 
$(\overline{\alpha}, \overline{A})$,
and evaluate the confidence interval through a least-square minimization
on the parameter $B$ in the relation
\begin{equation}
\log Y = \overline{\alpha} +\overline{A} \log X +B \log (1+z).
\label{eq:bfit}
\end{equation}
In particular, for a given grid of values of $\{B_i\}$, we search for the
minimum of the merit function
\begin{equation}
\chi^2_i = \sum_j \frac{\left[ \log Y_j -\overline{\alpha} -\overline{A} \log X_j
-B_i \log (1+z_j) \right]^2}{\epsilon_{\log Y_j}^2 
+\epsilon_{\overline{\alpha}}^2 +\overline{A}^2 \epsilon_{\log X_j}^2
+\epsilon_{\overline{A}}^2 \log^2 X_j}, 
\label{eq:chi2}
\end{equation}
where the errors on the best-fit local values, $\epsilon_{\overline{\alpha}}$ and
$\epsilon_{\overline{A}}$, are considered and propagated with the uncertainties,
$\epsilon_{\log X} = \epsilon_X/(X \, \ln10)$ and $\epsilon_{\log Y} =
\epsilon_Y/(Y \, \ln10)$, on the measured quantities.

The local and intermediate-redshift relations considered in this work
have been rescaled to the assumed cosmology and overdensity as
described in the Introduction. 
On the other hand, many results on the clusters scaling laws, 
and particularly on the $L-T$ relation, 
are presented in literature with 
the omission of the cosmological factor $E_z$ (that should be properly
considered only when the quantities are measured within a fixed overdensity
with respect to the critical density as done in the present work). 
Therefore, we have considered in the following analysis both the inclusion
and the omission of the factor $E_z$ to correct the quantities whose 
relations are investigated. 
In particular, we refer to the parameter $B$
for the evolution in the scaling relations not corrected by $E_z$ and
to the parameter $B_z$ for the evolution in the corrected scaling laws. 

It is worth noticing that the cosmological factor $E_z$ is defined as
$H_z / H_0 = \left[\Omega_{\rm m} (1+z)^3 + 1 - \Omega_{\rm m} \right]^{1/2}$
(for a flat cosmology with matter density $\Omega_{\rm m}$) and is
exactly equal to $(1+z)^{1.5}$ in an Einstein-de Sitter universe and
proportional to $(1+z)^{0.6/0.9}$ in the redshift range here considered
for an assumed $\Lambda$CDM model with $\Omega_{\rm m}=0.3$.\footnote{The
slope of the dependence of $E_z$ upon $(1+z)$ is the values
of the ratio $\log E_z / \log(1+z)$ for an assumed cosmology and a given
range in redshift.}
Furthermore, considering its dependence upon the redshift, this factor
affects significantly the quoted values of, e.g., mass and luminosity
for objects at high$-z$.
For example, the X-ray luminosity has to be corrected by a factor of
0.81 at $z=0.4$ and 0.51 at $z=1.2$, making these corrections fundamental
to proper calibrate the scaling relations in a sample of clusters
distributed over a wide interval in redshifts.
Thus, the omission of the factor $E_z$, that affects each quantity
in a different way at different redshift as stated above, has to be
properly considered once one wants to assess any evolution of the physical 
quantities that is not due to the expansion of the Universe 
and to the rescaling of the characteristic density.

With this notation in hand, we should observe $B_z=0$ and $B$ equal to the 
expected dependence of the scaling law upon $E_z$ if no evolution 
of the physical quantities is present. In the following analysis, only
the evolution of the $L-T$ relation is presented by separating $B$ and $B_z$.
One should bear in mind that the $L-T$ relation is then proportional to $E_z$
implying that $B$ should be measured in the range $0.6-0.9$. 

We describe below the constraints that we obtain on the parameters $B$ and 
$B_z$, whose $\chi^2$ distribution is plotted in the bottom panel of the figure 
that shows the corresponding scaling relation. All the results, obtained 
also by selecting different redshift-range in our sample, are presented
in Table~\ref{tab:bfit}.

\subsection{Evolution in the $L-T$ relation}

We compare the distribution of our measurements in
the $L-T$ plane (Fig.~\ref{fig:lt}) with the best-fit results obtained
for homogeneous and representative sets of nearby clusters from (i)
Markevitch (1998), who analyzed a sample of 31 clusters in the
redshift range 0.04--0.09 with measured \asca temperatures and \rosat
luminosities and excises systematically their inner 100 kpc to avoid
any contamination from central cooling cores, (ii) Arnaud \& Evrard
(1999) that include in their sample 24 clusters with weak or no
cooling flows in their cores and X-ray measurements from {\it GINGA},
\asca and {\it Einstein}, (iii) Novicki, Sornig \& Henry (2002, Tab.~4
and 5), that use a sample of 53 low ($z<0.3$) and 32 intermediate
($0.3<z<0.6$) redshift clusters observed with \asca.  
Considering that the best-fit results obtained from these samples do not 
include the correction by the cosmological factor $E_z$, 
we consider here in the same way our measurements without this correction
(Fig.~\ref{fig:lt}, panel on the left). In this case, the $L-T$ relation
should evolve like $E_z \approx (1+z)^{0.6/0.9}$ (for a ``$\Omega_{\rm m}=
1-\Omega_{\Lambda}=0.3$" universe) and we should measure $B=0.6-0.9$, 
if self-similar predictions are correct.

Moreover, we compare the distribution of our data corrected by $E_z$
with the cosmology-corrected relations obtained from resolved temperature
of (iv) 20 nearby $T>3$ keV
objects observed with \sax (Ettori, De Grandi \& Molendi 2002) and
(v) six clusters with redshift between 0.10 and 0.46 observed with
\chandra from Allen et al. (2001).
A predicted $E_z^{-1} L- T$ relation constant in time, and 
$B_z=0$, should be then observed.

When we compare the distribution of our data not corrected by $E_z$
with the local relations obtained from samples (i)--(iii) (Fig.~\ref{fig:lt}, 
panel on the left), we measure $B$ between $0.1$ and $0.6$.
In particular, when we use as local reference the result by Arnaud \& Evrard 
(1999), $B$ becomes significantly positive ($0.98 \pm 0.20$), but with a 
$\chi^2$ statistic ($\chi^2 = 51.6$, d.o.f. $= 27$)
that is worst by a $\Delta \chi^2 \ga 30$ than the others measurements,
mainly because of the relative small error on the slope $A$.
These measurements of $B$ are slightly lower than, but still consistent with,
the expected value between 0.6 and 0.9.
The evidence that $B$ tends to be lower then the predicted value seems to 
suggest the presence of a slight {\it negative} evolution, with clusters 
at higher redshift having lower bolometric luminosities for given
temperature.
However, the small departure from $B=0$ in our $L-T$ relation not corrected 
by $E_z$ agrees with previous determinations obtained from
Mushotzky \& Scharf (1997), Fairley et al. (2000) and Holden et al. (2002).
In particular, Holden et al. analyze a subsample of the present dataset
with only 12 objects, nine of those were exposed with \chandra.
On the other hand, this value of $B$ 
is significantly lower than the result presented in Vikhlinin et al. (2002).
These authors have assembled 22 clusters in the redshift range $0.39-1.26$ 
which were available with sufficient long exposure to ensure a good temperature
determination in the \chandra archive as of Spring 2002. 
Single temperatures (with column density and metallicity fixed to Galactic value 
and 0.3 times the solar abundance, respectively) were estimated collecting 
photons within 0.5--1 $h_{50}^{-1}$ Mpc. X-ray luminosities were measured
within 2 $h_{50}^{-1}$ Mpc. Both in the spectral and spatial analysis, the 
central regions with $r=$100 $h_{50}^{-1}$ kpc were excised if a sharply peaked
surface brightness was detected. 
Once compared with the local $L-T$ distribution 
in Markevitch (1998), where the same procedure in the determination
of temperatures and luminosities has been adopted, Vikhlinin et al. 
claim a detection significant at $8 \sigma$ confidence level 
of a positive evolution ($B=1.5 \pm 0.2$).

To check what causes this relevant difference, we
re-evaluate $B$ for the sample by Vikhlinin et al., using the results
by Markevitch (1998) for the low--$z$ reference relation. We apply to
this sample the same analysis procedure applied to our set of 28
clusters, which is based on minimizing the $\chi^2$ in
equation~\ref{eq:chi2} while propagating uncertainties both in
normalization and slope. We obtain $B=1.1 \pm 0.3$, with a deviation
from zero of only $3.5 \sigma$. 
This significance decreases to $2.2 \sigma$ ($B=0.62 \pm 0.28$,
$\chi^2=22.0$ with 27 d.o.f.) when our estimates of $L(<R_{500})$ 
(see asterisks in Fig.~\ref{fig:v02}) are considered for our sample 
of 28 objects.  
This value of $B$ is well in agreement with the predicted range of
$0.6-0.9$ expected when a face value of luminosity not corrected 
by $E_z^{-1}$ is considered.

Furthermore, it is worth noticing that the procedure adopted here defines
the radius, $R_{\Delta}$, at which all the physical properties of a system 
are recovered (see discussion in Sect.~2) as a function of its mass
(temperature), redshift and given overdensity. As we show in
Fig.~\ref{fig:rd}, the value of $R_{\Delta}$ varies considerably in
the redshift range 0.4--1.3 with respect to a fixed metric radius.
Therefore, systematic differences between quantities (such as X-ray
luminosity) measured at 2 $h_{50}^{-1}$ Mpc and $R_{500}$ are expected
and, in fact, are observed.
If we extrapolate further our estimates of luminosity up to $R_{200}$,
on average increase of 8 per cent in $L(<R_{200})$ with respect
to $L(<R_{500})$ is measured and a more significant evolution
($B \approx 0.7 \pm 0.3$) is observed. 

Therefore, we conclude that the mismatch between our and Vikhlinin et al. 
(2002) results is due to, in decreasing order of relevance, 
(i) the different fitting procedure (that reduces 
the significance of the evolution by propagating properly the errors on the 
best-fit parameters of the local relation), (ii) being our sample
larger at higher redshift (e.g., we have 16 galaxy clusters at $z>0.6$ and
4 at $z>1$, whereas Vikhlinin et al. have 9 and 2 objects, respectively),
(iii) systematic differences in the definition of the reference radius
and in the procedure to estimate the gas temperature and luminosity
(see also discussion at the end of Sect.~2.1 and Fig.~\ref{fig:v02}). 

We consider now the cosmological corrected $L-T$ relation, that should not
evolve in the self-similar scenario.
We compare our observed luminosities, which were corrected by the $E_z$ factor,
to the best-fit relations obtained from the samples (iv--v) discussed above.
We measure $B_z = -1.04 \pm0.32$ when our data are
compared to the determination by Ettori et al. (2002) at median
redshift of 0.05, and $B_z = -0.16\pm 0.23$ when comparing to the results by
Allen et al. (2001) for clusters at median redshift of 0.28
(Fig.~\ref{fig:lt}, panel on the right).
When we consider only the 16 objects with $z>0.6$, $B_z$ is measured more 
significantly negative: $B_z = -1.48 \pm 0.38$ and $-0.72 \pm 0.28$, respectively.
We also note that the $\chi^2$ is definitely improved (values between 
2 and 9 with 15 d.o.f.) also when the $B$ parameter is evaluated ($B$ values 
in the range $-0.5 \pm 0.7$ and $0 \pm 0.3$).

In agreement with a value $B$ lower than 0.6, a $B_z < 0$ indicates 
that a marginal detection of {\it negative} evolution
(i.e. clusters at higher redshift have lower luminosities 
for a given $T$) appears when the relation is properly
corrected by its cosmological dependence and compared to local 
estimates obtained under the same condition (i.e. all the photons
are considered to recover both temperature and luminosity and the $E_z$
factor is taken into account as done in Allen et al. 2001 and Ettori
et al. 2002).

\subsection{Evolution in the $M_{\rm tot}-T$ relation}

We compare our distribution in the $E_z M_{\rm tot}-T$ plane 
with the local relations measured by
(i) Ettori et al. (2002; \sax data), (ii) Finoguenov, Reiprich \& 
B\"ohringer (2001) for their subsample of 26 clusters with spatially
resolved temperatures and emission-weighted values larger than 3 keV
as measured with \asca (we have re-fitted their relation after rescaling 
the mass measurements in our cosmological model and correcting them
by the corresponding factor $E_z$) and (iii) Allen et al. (2001; \chandra data).
In Figure~\ref{fig:mt} we show our constraints on the evolution parameter
$B_z$.
Overall, there is no evidence of any evolution with values of $B_z$ 
always consistent with zero: $B_z = -0.12 \pm 0.39$ (Ettori et al. 2002),
$B_z = -0.26 \pm 0.17$ (Finoguenov et al. 2001) and
$B_z = 0.02 \pm 0.39$ (Allen et al. 2001).
Also if we restrict the measurements to the subsample of clusters with $z>0.6$, 
we obtain $B_z$ consistent with zero: $B_z = -0.14 \pm 0.46$ (Ettori et al. 2002),
$B_z = -0.16\pm0.21$ (Finoguenov et al. 2001) and
$B_z = 0.04\pm0.45$ (Allen et al. 2001).

\subsection{Evolution in the $M_{\rm gas}-T$ relation}

Fig.~\ref{fig:mgt} shows that there is no significant evolution in the 
$M_{\rm gas}-T$ relation corrected by $E_z$ when it is
compared with local estimates from Mohr et al. (1999) and Ettori et
al. (2002). For this relation, we find $B_z = -0.10\pm0.20$ and
$-0.26\pm0.23$ for these two reference analyses, thus corresponding to
a detection of a negative evolution at $\la 1 \sigma$ confidence level
(in case of no evolution, $B_z =0$).
A slightly more significant evolution is detected when $z>0.6$ clusters
are used: $B_z = -0.42 \pm 0.24$ and $-0.54 \pm 0.27$, respectively. 
Still, it is worth noticing that, as observed in
the $L-T$ relation, if any hint of evolution can be detected, this
points toward a negative evolution, i.e. clusters tend to have lower
X-ray luminosity and gas mass at higher redshift for given
temperature. We underline that this is only a marginal ($<2 \sigma$)
evidence, even if confirmed from both measurements of X--ray
luminosity and gas mass.

\subsection{Evolution in the $L-M_{\rm tot}$ relation}

The cosmologically corrected $L-M_{\rm tot}$ relation (Fig.~\ref{fig:lm})
appears consistent with a case of slightly negative evolution
(i.e. $B_z < 0$) that affects particularly the X-ray luminosities at
redshift greater then 0.6:
we measure $B_z = -0.60\pm0.17$ ($-1.10\pm0.20$ for objects with $z>0.6$)
and $0.18\pm0.20$ ($-0.86\pm0.25$ at $z>0.6$) when compared with
the results in Ettori et al. (2002) and Reiprich \& B\"ohringer (2002),
respectively. 
Note that we have obtained and considered the best-fit constraints on 
the $E_z^{-1} L_{\rm bol} - E_z M_{500}$ relation for the 85 objects 
(out of 106) in the Reiprich \& B\"ohringer (2002) sample with quoted 
X-ray temperature larger than 3 keV. 
Even though the scatter that we measure in this relation is the highest among
the scaling laws investigated, the trend observed for systems with lower
luminosities for given mass at higher redshifts agrees with the results
presented on the $L-T$ relation.

\subsection{Evolution in the $S-T$ relation}

\begin{figure}
\epsfig{figure=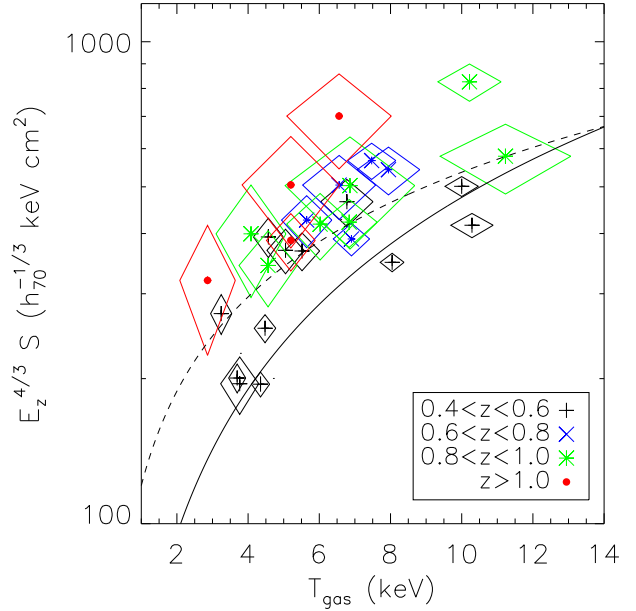,width=0.5\textwidth}
\caption{Distribution of the entropy values measured at $0.1 \times R_{200}$
versus gas temperature. The distribution shows a Spearman's rank 
correlation significant at 99.9 per cent against the null-hypothesis that
no-correlation is present.
A rescaling factor of $E_z^{4/3}$ is used to compare these values
to what is expected from hydrodynamical simulations, $S \approx 50 \ 
(f_{\rm gas}/0.106)^{-2/3} \ T_{\rm keV} \ h_{70}^{-4/3}$ kev cm$^2$
(solid line, from Ponman et al. 1999 and normalized to the median gas fraction
measured in our sample; dashed line, $S \approx 120  \ T_{\rm keV}^{0.65}$, from
Ponman et al. 2003).
} \label{fig:entr} \end{figure}

As a further diagnostic to trace the evolution of the ICM properties,
we analyze the relation between entropy, $S$, and gas temperature.
Starting from the work by Ponman et al. (1999), entropy has received a
considerable attention as a diagnostic of the physical processes that
establish the thermal status of the diffuse baryons in clusters (e.g.,
Tozzi \& Norman 2001, Voit et al. 2002). For the first time, we
attempt here to trace the evolution of the ICM entropy, whose analysis
has been limited so far to low--$z$ clusters.  To this purpose, we
measure the quantity $S= T_{\rm gas} / n_{\rm e}^{2/3}$ at $0.1 \times
R_{200}$ within each cluster 
as done by Ponman and collaborators for nearby systems.
With the above definition, the quantity
$S$ relates to the thermodynamical entropy $K$ of the system as $K
\propto \log S$. 

In Figure~\ref{fig:entr}, we plot the values of $S$ as a function of
the gas temperature. The predicted scaling of $S \sim T$ is in
agreement with the overall fit of the data [$S = (407 \pm 20) \times
T_{\rm gas}^{0.99\pm0.20}$ keV cm$^2$].  A mild negative evolution ($B_z
=-0.14 \pm 0.04$, $\chi^2_{\rm min} = 151$ with 27 degrees of freedom)
is observed when all our sample is compared with the {\it local}
estimate of the ``entropy ramp'' by Ponman et al. (2003; see
Fig.~\ref{fig:entr}, panel on the left).  However, the high $\chi^2$
value warns that this is not a good modelization of the data
distribution, mainly for their large internal scatter.  If we consider
only the objects at $z> 0.6, 0.8$ and $1$, respectively, we observe
consistently $B_z \approx 0.3$ with a deviation from no evolution 
($B_z =0$) at the
$2-3 \sigma$ confidence level, with a reduced $\chi^2$ of about 1
(e.g., $B_z=0.26 \pm 0.06, \chi^2$/d.o.f. $=18.5/15$ at $z> 0.6$, 
$B_z=0.32 \pm 0.08, \chi^2$/d.o.f. $=10.6/10$ at $z> 0.8$).
%

\subsection{Discussion on the observed evolution}

As for the $E_z^{-1} L-T$ and $E_z M_{\rm gas}-T$ relations, we detect a mild
negative evolution [$B_z \sim (-0.1, -1.0)$, with a deviation from the
non-evolution case, $B_z=0$, by less than $3.2 \sigma$] that becomes 
more relevant when subsamples with higher cut in redshift are considered.
For the 16 objects with $z>0.6$, we measure $B_z$ in the range $[-1.5, -0.4]$ 
with deviation from zero significant at $3.9 \sigma$ (see Table~\ref{tab:bfit}).
This result points to a scenario in which clusters at higher
redshift have in general lower X-ray luminosities and gas masses than
what predicted from the self-similar model.
This trend appears more clearly when the entropy of these hot ($T_{\rm
gas} >$ 3 keV) systems, evaluated at $0.1 R_{200}$, is compared with
local determinations.  A mild {\it positive} evolution ($B_z \simeq
0.3$) is then detected in the $E_z^{4/3} S-T$ relation, in particular 
when objects at redshift larger than 0.6 are considered.

The evolution of the $L-T$ and $M_{\rm gas}-T$
relations is a sensitive test for non-gravitational heating models
(e.g. Tozzi \& Norman 2001, Bialek et al. 2001) and carries
information about the source of this heating and the typical redshift
at which it took place. The numerical simulations by Bialek et
al. (2001) suggest that $M_{\rm gas}$ is lower at higher redshift for
fixed temperature, when the ICM is heated to an initial entropy of
about 100 keV cm$^2$, well before the cluster collapse takes
place. This effect is more evident for smaller systems, thus making
the $M_{\rm gas}-T$ relation steeper at increasing redshift. A
marginal, negative evolution in the $L-T$ relation is predicted by the
semi-analytical model by Tozzi \& Norman (2001), when an
epoch-independent entropy background of $K = 0.3 \times 10^{34}$ ergs
cm$^2$ g$^{-5/3}$ is combined with cooling in a $\Lambda$CDM universe
(see their Fig.11 for emission-weighted temperature above 2 keV).  We
conclude that the observed negative evolution in the $L-T$ and $M_{\rm
gas}-T$ relations are in agreement with predictions from models that
pre-heat the ICM with a non-evolving entropy floor at the level of
about 100 keV cm$^2$.  In any case, this picture has been recently 
questioned by Ponman et al. (2003), who find no evidence for isentropic 
cores in groups.

A more controversial point is whether pre--heating can also account
for the positive evolution of entropy in central cluster regions. 
If we are looking just at the effect of an entropy floor, which has been
previously created by an extra--heating mechanism, 
the central regions of clusters would be characterized by an
entropy level which does not change with time.  Therefore, after
applying the cosmological rescaling, the quantity $E_z^{4/3}S$ should
be positively evolving. However, this evolution turns out to be stronger 
than that measured from the clusters at $z>0.6$, $(1+z)^{0.3}$, thus 
excluding that our measured evolution is the effect of a $z$--independent 
entropy floor. Furthermore, since most of our clusters are quite hot systems, 
they should sample the regime where entropy is dominated by gravitational 
shocks, rather than by a pre--collapse entropy floor.
Indeed, in the standard scenario of pre--heating, one expects 
a minimum entropy level to be created in central regions of groups and poor
clusters, while the ICM entropy structure should be left almost
unaffected for hot systems, like those included in our sample. 
More in general, the emerging scenario from the entropy
properties of low-- and high--$z$ systems is calling for alternative
interpretations, such as pre--heating with differential entropy
amplification, as recently proposed by Voit et al. (2003), or a
suitable combination of radiative cooling and heating by feedback
energy release (e.g., Tornatore et al. 2003; Kay et al. 2003).

A further possibility is that we are looking at the combined effect of
non--gravitational heating and radiative cooling. Although each one of
these two mechanisms provides an excess entropy, hydrodynamical
simulations of clusters have shown that their combined effect may
induce non--trivial consequences in the entropy properties of the ICM
(Tornatore et al. 2003). For instance, an episode of impulsive heating
may cause the creation on an excess entropy, thus increasing the gas
cooling time and suppressing cooling in central cluster
regions. Subsequently, cooling starts acting again on the gas,
gradually decreasing its entropy as gas flows toward the cluster
center. In this case, one should observe a gradual reduction of the
net entropy detected at the cluster center. Whether this is what we
are actually observing requires a more sophistical numerical and
semi--analytical modeling. In any case, there is little doubt that the
evolution of gas entropy in central cluster regions keeps track of the
past ICM thermal evolution. As such, its measurement provides unique
information about the latest epoch at which a significant amount of
energy has been injected into the ICM and of the involved energy
budget.

\begin{figure}
\epsfig{figure=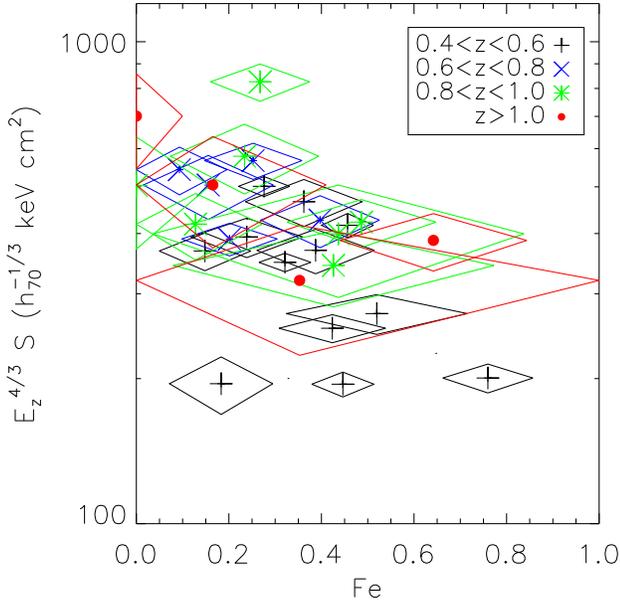,width=0.5\textwidth}
\caption{Distribution of the entropy values measured at $0.1 \times R_{200}$
versus the Iron abundance measured with respect to the solar ratios in 
Anders \& Grevesse (1989). The Spearman's rank correlation is significant at
99.0 per cent against the no-correlation case.
} \label{fig:entr_Z} \end{figure}

As a diagnostic to connect the thermodynamics of the ICM to the past
history of star formation, we have investigated the relation between
entropy and the Iron content. (We remind that in
these high$-z$ systems Iron is the only element that we can sample in
their metal budget.) As shown in 
Fig.~\ref{fig:entr_Z}, we have a statistically significant indication
that values of the central entropies are lower in correspondence to
metal-richer clusters. This result, along with the evidence that
systems at lower temperature stores lower amount of entropy at the
same fraction of the virial radius, is an alternative way to show what 
we reported in a previous paper (Tozzi et al. 2003): clusters at lower
temperature appear to have higher Iron abundance (see Baumgartner et al. 2003
for a discussion of this trend as observed in the \asca galaxy 
cluster archive).

It is now well-established (e.g. Pettini 2003, Finoguenov et al. 2002)
that most of the star formation and Iron production happened at $z \ga
2$, i.e. before the final collapse of the accreting baryons into
clusters, and preferentially in low-entropy structures like groups
along cosmic filaments. Groups are the high--$z$ precursor of
later--forming rich clusters, which maintain their structure and
survive for a few crossing times after they merge into the cluster
environment (as suggested by the frequently observed {\it cool cores}
in relaxed systems -e.g. Vikhlinin et al. 2001- and corroborated by
hydrodynamical simulations -Motl et al. 2003, and Tormen et al. 2003). 
This implies that the shorter
the time elapsed from the aggregation of these clumps in a larger
system, the larger the number of survival regions at low entropy and
high metallicity. It is worth noticing that these stable subclusters
should contribute significantly to all the X-ray emission-weighted
observables because of their high gas density. Therefore, from an
observational point of view, systems with higher metallicity should
appear with lower gas temperature and be considered as dynamically
younger. This supports what is shown by the surface brightness
profiles, with centrally peaked emission, which is characteristic
of relaxed systems, found to be less evident as more distant
clusters are considered.

\section{Summary and Conclusions}

We have studied the evolution of the scaling relations between
observed and derived physical properties of the ICM and dark matter
potential in a sample of 28 X-ray galaxy clusters with redshift in the
range $z\simeq 0.4-1.3$ (median of $0.73$).  Using \chandra
observations, we resolve their surface brightness on scales of a few
arcsecs (1 arcsec corresponds to between 5 and 8 $h_{70}^{-1}$ kpc at
these redshifts for the assumed cosmology), out to a significant
fraction (about 0.7, as median value) of $R_{500}$. 

In general, the surface brightness is well fitted by a single
$\beta-$model. A double $\beta-$model is statistically not required,
suggesting that these distant clusters do not show any significant
central brightness excess. This may be indicative that systems at high
redshift appear as structures in formation and not completely relaxed.
By collecting all the photons within $\sim$0.5 $\times R_{500}$, we
have measured a single emission-weighted temperature (median value of
6.0 keV). The combination of the spatial and spectral analysis allows 
us to recover the gas luminosity (observed in the range
$1.0-116.7 \times 10^{44} h_{70}^{-2}$ erg s$^{-1}$), 
gas mass ($0.6-18.1 \times 10^{13} h_{70}^{-5/2} M_{\odot}$) 
and total gravitating mass ($1.4-21.3 \times 10^{14} h_{70}^{-1} M_{\odot}$) 
out to $R_{500}$. These quantities are used to quantify the evolution of 
the X--ray scaling relations in hot ($T_{\rm gas} \ga 3$ keV) clusters.

The slopes of all the investigated correlations tend to be higher than
what is predicted by self-similar models. The largest deviations from
the self-similar scaling relations occur
in the $L-T$ and $M_{\rm gas}-T$ relations (Fig.~\ref{fig:lt},
\ref{fig:mt}, \ref{fig:mgt}). 
We notice that the values of the slope
propagate through the relations in a self-consistent way: for example,
the observed dependence $L \sim M_{\rm tot}^{1.88 \pm 0.42}$ and 
$M_{\rm tot} \sim T^{1.98 \pm 0.30}$ should imply $L \sim T^{3.7 \pm 1.0}$
that is well in accordance with the measured slope of $3.72 \pm 0.47$.
This is in accordance
with the observed properties of gas-dependent quantities, such as
luminosity and gas mass (e.g. Edge \& Stewart 1991, Fabian et
al. 1994, Mushotzky \& Scharf 1997, Allen \& Fabian 1998, Mohr et
al. 1999, Novicki et al. 2001, Ettori et al. 2002), and suggests that
simple gravitational collapse is not the only process that governs the
heating of baryons in the potential well of clusters also in the redshift
range $0.4-1.3$.
On the other hand, we do not observe any significant dependence of the 
gas mass fraction upon the temperature (see Fig.~\ref{fig:ltm}), ruling
out the simple explanation that the increase in the gas budget
as function of the mass of the system is responsible for the 
steepening of the $L-T$ relation.  

To explain the observed correlations, a raise in the central entropy
(and a corresponding suppression of the X-ray emissivity) is required
and can be achieved either by episodes of non-gravitational heating
due to supernovae and AGN (e.g., Evrard \& Henry 1991; Cavaliere,
Menci \& Tozzi 1999; Tozzi \& Norman 2001; Bialek, Evrard \& Mohr
2001; Brighenti \& Mathews 2001; Babul et al. 2002; Borgani et
al. 2002), or by selective removal of low-entropy gas through cooling
(e.g. Pearce et al. 2001; Voit \& Bryan 2001; Wu \& Xue 2002),
possibly regulated by some mechanism supplying energy feedback (e.g.
the semi-analytical approach proposed by Voit et al. 2002 and the
numerical simulations discussed by Muanwong et al. 2002, Tornatore et
al. 2003, Kay et al. 2003).

When the slope of the $E_z M_{\rm tot}-T$ relation is fixed to 1.5, its
high--$z$ normalization of $(0.37 \pm 0.02) \times 10^{14} h_{70}^{-1} 
M_{\odot}$ keV$^{-3/2}$ is well in agreement with results from low--$z$ 
samples (e.g. Nevalainen et al. 2000, Allen et al. 2001). This lack 
of evolution follows from the prediction of hydrostatic equilibrium 
within virialized halos. In turn, this is consistent with
the expectation that the $E_z M_{\rm tot}$--$T$ relation is the least
sensitive to the thermal status of the gas and, therefore, not
significantly affected by non--gravitational processes. Still, the
$M_{\rm tot}$--$T$ relation observed at low redshift (e.g., Horner et
al. 1999, Nevalainen et al. 2000, Finoguenov et al. 2001, Ettori et
al. 2002) is known to have a normalization lower by about 30 per cent
than that obtained from hydrodynamical simulations (e.g. Mathiesen \&
Evrard 2001, cf. also Muanwong et al. 2002, Borgani et al. in
preparation). It will be interesting to show how our measured $E_z M_{\rm
tot}$--$T$ for distant clusters compare with predictions from
hydrodynamical simulations.

 
As for the $E_z^{-1} L-T$ and $E_z M_{\rm gas}-T$ relations, we detect a mild
negative evolution [$B_z \sim (-0.1, -1.0)$, with a deviation from the
non-evolution case $B_z=0$ by less than $3.2 \sigma$] 
that becomes more relevant when subsamples with higher cut in redshift are 
considered (see Table~\ref{tab:bfit}).
For the 16 objects with $z>0.6$, we measure $B_z$ in the range $[-1.5,-0.4]$ 
with deviation from zero significant at $3.9 \sigma$.
This result points to a scenario in which clusters at higher
redshift have in general lower X-ray luminosities and gas masses than
that predicted from the self-similar model.
This trend appears more clearly when the entropy of these hot ($T_{\rm
gas} \ga 3$ keV) systems, evaluated at $0.1 R_{200}$, is compared with
local determinations.  A mild {\it positive} evolution ($B_z \simeq
0.3$) is then detected in particular when objects at redshift larger
than 0.6 are considered.
The cut at $z \ge 0.6$ allows us to improve
significantly the fit when a $\chi^2$ statistic is applied to evaluate
the level of evolution. Even though a power-law modelization is probably
simplistic to represent the properties of the ICM as a function of redshift, 
it provides generally a very good fit when clusters
with $z<0.6$ are not included. 
 
As we discuss in Section~4.6, these observed evolutions are broadly 
consistent with predictions from models that
pre-heat the ICM with a non-evolving entropy floor at the level of
about 100 keV cm$^2$. However, these
models are in conflict with recent observations of non-isentropic
cores in galaxy groups (e.g. Ponman et al. 2003, Mushotzky et al. 2003).
More detailed analyses of simulated structure formation are required to 
confirm quantitatively the degree of the expected evolution. 

\section*{Acknowledgements}
We thank the anonymous referee for a careful reading of the manuscripts 
and suggestions that have improved the presentation of our work.
We are grateful to Thomas Reiprich for his quick reply to our emails.

\newpage
\appendix
  \renewcommand{\thefigure}{A-\arabic{figure}}
  \setcounter{figure}{0}  

\begin{figure*}
\input{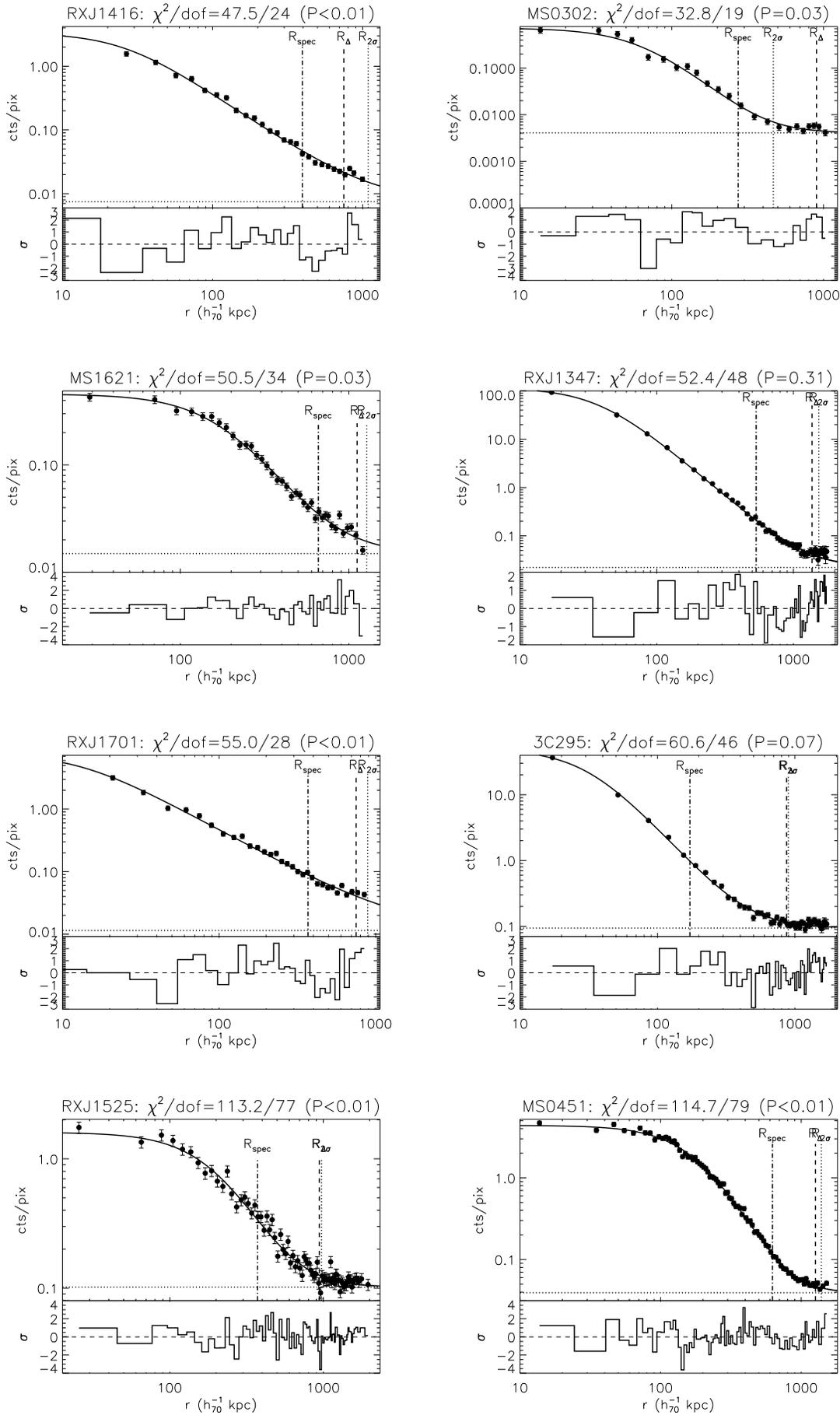}
\caption{Data, best-fit $\beta$-model and residuals
of the surface brightness profile 
of the high redshift galaxy clusters in our sample.
Dotted lines indicate the best-fit background value.
The location of $R_{\rm spec}$, $R_{\Delta}$ and
$R_{2\sigma}$ is shown.
}
\end{figure*}

\begin{figure*}
\input{fig_app_2.tex}
\caption{(Continue) }
\end{figure*}
\begin{figure*}
\input{fig_app_3.tex}
\caption{(Continue) }
\end{figure*}
\begin{figure*}
\input{fig_app_4.tex}
\caption{(Continue) }
\end{figure*}

\end{document}